\documentclass[12pt,preprint]{aastex}
\usepackage{color}
\usepackage{amsbsy}

\newcommand{\vc}[1]{{\boldsymbol #1}}
\newcommand{\de}{\mathrm{d}}
\newcommand{\dpa}{\partial}
\newcommand{\De}{\mathrm{D}}
\newcommand{\nab}{\vc{\nabla}}
\newcommand{\ee}{\mathrm e}
\DeclareMathSymbol{\varOmega}{\mathord}{letters}{"0A}
\DeclareMathSymbol{\varSigma}{\mathord}{letters}{"06}
\DeclareMathSymbol{\varPsi}{\mathord}{letters}{"09}

\newcommand{\Eq}[1]{equation (\ref{#1})}
\newcommand{\Eqs}[2]{equations (\ref{#1}) and~(\ref{#2})}
\newcommand{\App}[1]{Appendix~\ref{#1}}
\newcommand{\Sec}[1]{Sect.~\ref{#1}}

\newcommand{\Fig}[1]{Fig.~\ref{#1}}
\newcommand{\Figs}[2]{Figs.~\ref{#1} and \ref{#2}}
\newcommand{\Tab}[1]{Table \ref{#1}}

\shorttitle{Dust Diffusion by Magnetorotational Turbulence}
\shortauthors{Johansen \& Klahr}

\begin{document}


\title{Dust Diffusion in Protoplanetary Discs by Magnetorotational Turbulence}


\author{Anders Johansen\altaffilmark{1} and Hubert Klahr}
\affil{Max-Planck-Institut f\"ur Astronomie, K\"onigstuhl 17, 
    69117 Heidelberg, Germany}
\email{johansen@mpia.de}


\altaffiltext{1}{also at: NORDITA, Copenhagen, Denmark}


\begin{abstract}

We measure the turbulent diffusion coefficient of dust grains embedded in
magnetorotational turbulence in a protoplanetary disc directly from numerical
simulations and compare it to the turbulent viscosity of the flow. The
simulations are done in a local coordinate frame comoving with the gas in
Keplerian rotation. Periodic boundary conditions are used in all directions,
and vertical gravity is not applied to the gas. Using a two-fluid approach,
small dust grains of various sizes (with friction times up to $\varOmega_0
\tau_{\rm f}=0.02$) are allowed to move under the influence of friction with
the turbulent gas. We measure the turbulent diffusion coefficient of the dust
grains by applying an external sinusoidal force field acting in the vertical
direction on the dust component only. This concentrates the dust around the
mid-plane of the disc, and an equilibrium distribution of the dust density is
achieved when the vertical settling is counteracted by the turbulent diffusion
away from the mid-plane. Comparing with analytical expressions for the
equilibrium concentration we deduce the vertical turbulent diffusion
coefficient. The vertical diffusion coefficient is found to be lower than the
turbulent viscosity and to have an associated vertical diffusion Prandtl number
of about 1.5. A similar radial force field also allows us to measure the radial
turbulent diffusion coefficient. We find a radial diffusion Prandtl
number of about 0.85 and also find that the radial turbulent diffusion coefficient is
around 70\% higher than the vertical. As most angular momentum transport
happens through magnetic Maxwell stresses, both the vertical and the radial
diffusion coefficients are found to be significantly higher than suggested by
the angular momentum transport by Reynolds stresses alone. We also find
evidence for trapping of dust grains of intermediate friction time in turbulent
eddies.

\end{abstract}




\verb|$Id: ms.tex,v 1.4 2005/08/17 07:59:04 ajohan Exp $|

\section{INTRODUCTION}

Knowledge of the transport properties of particles embedded in a turbulent gas
medium is important in many aspects of protoplanetary disc modeling. If
the spatial number density distribution of dust grains in a disc is required
for the model, one must know the effect of turbulent diffusion on the dust
grains.

{\em Vertical diffusion} --- The distribution of tiny dust grains, with radii
smaller than around 100 $\mu$m, determines the observability of protoplanetary
discs around young stellar objects through their contribution to the infrared
parts of the spectrum. An interesting observational effect of turbulent
diffusion is its influence on the vertical settling of dust grains. The
settling affects the spectral energy distribution of protoplanetary discs,
since flaring discs, i.e.\ where the scale height of the gas density increases
with radial distance, have a much stronger mid- to far-infrared excess than
self-shadowing discs, where the scale height after a certain distance from the
protostar begins to fall with radial distance \citep[e.g.][]{Dullemond2002}.
Recent model calculations by \cite{DullemondDominik2004} show that the vertical
settling of dust grains towards the mid-plane of the disc can change an
initially flaring disc into a partially self-shadowing disc, thus effecting the
observability of the disc. These calculations depend -- among other things -- on the
strength of the turbulence in the disc (the turbulent viscosity) and on the
turbulent diffusion coefficient of dust grains in the direction perpendicular
to the disc mid-plane. Also, \cite{Ilgner+etal2004} recently considered the
effect of vertical mixing in protoplanetary discs on the distribution of
various chemical species and found the distribution to be influenced greatly by
mass transport processes, again underlining the importance of vertical
turbulent diffusion in the modeling of protoplanetary discs.

{\em Radial diffusion} --- Crystalline silicate dust grain features observed in
comet spectra are often attributed to radial mixing in the solar nebula
\citep[e.g.][]{Hanner1999}. Silicate dust grains are formed primarily in
amorphous form, but they can become crystalline if exposed to temperatures
above $\sim1000$ K. Such a heating can obviously have occurred in the hot inner
parts of the solar nebula, whereas comets are expected to have formed in the
cold outer regions of the nebula, so in this picture some radial mixing must
take place between the inner and outer nebula. \cite{Bockelee-Morvan+etal2002}
consider disc models where crystallization of silicates happens in the inner,
hot parts of the disc. They calculate that in a few times $10^4$ years the
crystalline silicate fraction reaches a uniform value outside the
crystallization region due to radial turbulent diffusion, and that the value can
approach unity for realistic disc parameters. From high resolution
observations of three protoplanetary discs, \cite{vanBoekel+etal2004} find that
the inner 1-2 AU of these discs contain a higher crystalline silicate fraction
than the outer 2-20 AU. This supports the theory that crystalline dust grains form
in the hot inner disc and are subsequently transported to the outer disc by
turbulent gas motion.

The existence of chondrules \citep[millimeter-sized solid inclusions found in
primitive meteorites, see e.g.][]{Norton2002} is believed to be the result of
collisions and coagulation of small dust grains \citep{BlumWurm2000}. The size
distribution of chondrules may be explained by selective sorting in the
turbulent solar nebula \citep{Cuzzi+etal2001}. The first step in planet
formation is the build-up of kilometer-sized rocky and icy planetesimals
\citep[in the planetesimal hypothesis of][]{Safronov1969}, either from sticking
or due to a gravitational instability in the vertically settled dust layer. In
the latter case, the equilibrium scale height of the dust layer is determined by
the turbulent diffusion coefficient of the dust grains in the vertical
direction \citep{Cuzzi+etal1993}. An alternative planet formation hypothesis,
the gravitational instability hypothesis \citep[see][and references
therein]{Boss2003}, states that planets form as a direct gravitational
instability in the gas of a protoplanetary disc. The ability of a disc to
undergo gravitational instability depends on its density and temperature
structure, which is again dependent on the distribution and thus the turbulent
transport of tiny dust grains.

It is a modern paradigm of protoplanetary discs that shear instabilities in the
gas flow lead to turbulence, which is again responsible for such diverse
effects as heating, angular momentum transport and diffusion. The actual
turbulence is often parametrized in a single parameter, the turbulent viscosity
\citep[which can be non-dimensionalized into the $\alpha$-value
of][]{ShakuraSunyaev1973}. This single parameter determines both heating,
angular momentum transport and diffusion. Candidates for protoplanetary disc
turbulence are many. Most pronounced linear instabilities are vertical
convective instability \citep{LinPapaloizou1980} and the magnetorotational
instability \citep{BalbusHawley1991}, although the former has proved to lead to
inward rather than outward transport of angular momentum 
\citep{RyuGoodman1992}. Other instabilities have been proposed, such as the
baroclinic instability of \cite{KlahrBodenheimer2003} which must be non-linear
according to \cite{Klahr2004}, a linear Rossby wave instability
\citep{Li+etal2000} and a linear stratorotational instability
\citep{Dubrulle+etal2005,ShalybkovRuediger2005}.

Today's most accepted source of turbulence is magnetorotational turbulence
(MRI). For
completely ionized discs, the emergence of self-sustained turbulence through
the linear magnetorotational shear instability seems inevitable, both in local
shearing box simulations \citep{Brandenburg+etal1995,Hawley+etal1995} and in
global accretion disc simulations \citep{Armitage1998,ArltRudiger2001}. The
application of the ideal MHD equations to protoplanetary discs is only
justified where the ionization fraction is relatively high
\citep[e.g.][]{Fromang+etal2002,Semenov+etal2004}. This may be given in the hot
and dust-free inner parts of the disc, as well as away from the mid-plane of the
disc and at large radial distances where the ionization is determined by cosmic
ray and high energy photon penetration. In protoplanetary discs this had lead
to the concept of a magnetically dead zone near the mid-plane of the disc where
the ionization fraction is too low to sustain MRI. \cite{FlemingStone2003}
consider local shearing box simulations with a vertically dependent ionization
fraction and find that some turbulent stresses are induced in the dead
zone by the surrounding MRI turbulence. Thus angular momentum can be
transported even in regions of the disc that are not magnetorotationally
unstable.

It is often assumed that turbulent transport takes place as diffusion. For dust
grains, the turbulent flux is assumed proportional to the gradient of the
dust-to-gas ratio \citep{Dubrulle+etal1995}. Such a prescription does not per
se determine a certain value for the turbulent diffusion coefficient. Hence
it 
is often parametrized to be a scalar that is equal to the turbulent viscosity
of the gas for tiny grains but falls gradually for larger and larger grain
sizes \citep{Cuzzi+etal1993,SchraeplerHenning2004}. One argument for setting
the turbulent diffusion coefficient equal to the turbulent viscosity is that
the radial velocity fluctuations are the base of both (non-magnetic) angular
momentum transport and diffusion \citep[][p. 143]{TennekesLumley1972}. Such an
approach is simple to use, but caution should be taken regarding its validity,
both regarding the numerical value of the turbulent diffusion coefficient and
regarding the isotropy that is implicitly assumed when making it a scalar. 

The validity of the whole diffusion description must also be addressed
An obvious cause of concern is the presence of
dust-trapping mechanisms in the turbulent gas flow. Gas turbulence and global
pressure gradients, e.g.\ from vertical and radial stratification, are the
cause of two important trapping mechanisms. Whenever the gas is
pressure-supported and in force balance, the embedded dust grains feel an
excess force in the opposite direction to the gas pressure gradient, since they
can never be in the same force equilibrium without pressure support. The dust
grains thus feel an acceleration relative to the gas. This has various effects, e.g.\
vertical settling of the dust layer towards the mid-plane of protoplanetary
discs or inward radial drift of dust grains if there
is an outwards decreasing gas pressure in the disc, a notorious problem in
planet formation \citep{Weidenschilling1977}. The dust grains reach a terminal
velocity when the friction with the gas balances out the acceleration due to
the missing pressure gradient. The terminal velocity of very small dust grains
is proportional to the friction time. In non-turbulent disc models only global
pressure gradients are present, but in a turbulent disc local, fluctuating
regions of high and low pressure are expected to occur. Then dust grains
continuously move up the local pressure gradient, and this contributes to the
random motion of the grains, which is responsible for turbulent diffusion.
Magnetic pressure gradients actually give the same effect, as we will show
analytically in \Sec{ch:shorttaufappr}. A local concentration of dust grains
can not be described as diffusion, so one of the goals of this paper is to test
the validity of the global diffusion picture in the presence of turbulent
pressure gradient trapping.

For larger dust grains, where the friction time becomes comparable to the
orbital period of the disc, another important dust-trapping mechanism sets in.
Stationary rotational structures in the gas (e.g.\ anticyclones) are given by
an equilibrium between the global Coriolis force from the rotating disc and the
centrifugal force of the rotating structure. As they enter such a structure, large dust grains experience a
slow acceleration, due to drag forces with the gas.
Rotating initially with the gas, but much slower, the Coriolis force
dominates over the centrifugal force, and the dust grains are sucked into the
eddy. This vortex trapping was proposed by \cite{BargeSommeria1995}, and has
since then been subject of much theoretical investigation
\citep[e.g.][]{Chavanis2000,Johansen+etal2004}. The conclusions are that
vortices are extremely efficient at trapping dust grains, and this efficiency
may even explain how gas planets are formed before the dispersion of the gas
disc \citep{KlahrBodenheimer2005}. Vortex trapping would seem to be potentially
even more threatening to the global diffusion description than pressure
gradient trapping.

In this paper we measure the turbulent diffusion coefficient of dust grains
directly from numerical simulations of three-dimensional magnetorotational
turbulence. We treat the physics of protoplanetary discs in the shearing sheet
approximation, in which a local coordinate frame corotating with the disc is
considered. Dust is added as an extra fluid that interacts with the gas
through a drag force. The turbulent diffusion coefficient is measured by
exposing the dust fluid to an external force field and then comparing the
resulting dust density with analytical expressions derived with a parametrized
diffusion term. By comparing the measured value to the turbulent viscosity we
examine whether the two are indeed equal as is often assumed. We specifically
address the question of whether the diffusion coefficient is isotropic by
measuring diffusion in both the vertical and the radial direction. Finally we
quantify the effect of dust-trapping mechanisms on the whole diffusion picture
by examining correlations between turbulent gas features and the dust-to-gas
ratio. 

The paper is built up as follows. In \Sec{ch:dynamical_equations} we describe
the dynamical equations for the motion of gas and dust and the computer code
that we use to solve them numerically. Then we go into details in
\Sec{ch:diffusion_coefficient} about how we deduce the turbulent diffusion
coefficient from computer simulations by comparing the equilibrium dust density
with analytical expressions. In \Sec{ch:units} we describe the units and the
boundary conditions of the simulations. The results are described in the
following two sections; \Sec{ch:evolution_gas} describes the turbulent
evolution of the gas while \Sec{ch:evolution_dust} describes the evolution of
the dust, especially the measured turbulent diffusion coefficients and
diffusion Prandtl numbers and the presence of dust-trapping mechanisms in the
gas. Finally conclusions, discussions and some outlook to potential further
investigations into the subject of turbulent diffusion of dust grains appear in
\Sec{ch:conclusions}.

\section{DYNAMICAL EQUATIONS}
\label{ch:dynamical_equations}

In this section we present the dynamical equations we use for gas velocity, gas
density, magnetic vector potential, dust velocity and dust density. We
integrate the dynamical equations using the Pencil Code\footnote{The code is
available at\\\url{http://www.nordita.dk/software/pencil-code}}. This is
a finite difference code that uses sixth order centered spatial derivatives and
a third order Runge-Kutta time-stepping scheme. See \cite{Brandenburg2003} for
details on the numerical schemes and test runs. The Pencil Code solves the dynamical
equations in their non-conservative form and gives very similar
results to the ZEUS code for the statistical properties of MRI turbulence
\citep[see][and references therein]{BalbusHawley1998}.

The Pencil Code requires artificial diffusivity terms in the dynamical
equations to stabilize the finite difference numerical scheme. For the purpose
of calculating turbulent diffusion coefficients it is vital that we can reduce
the artificial mass diffusion as much as possible in order to distinguish the
measured turbulent diffusion from the imposed artificial diffusion. The biggest
contribution to the turbulent transport of dust grains is expected to come from
the fast and far moving large scales, so keeping the large scales unaffected by
diffusion is important. To achieve this we use hyperdiffusivity terms in all
the dynamical equations. Hyperdiffusivity involves replacing the regular
diffusivity terms (involving second order derivatives) with differential
operators that use higher order derivatives. This quenches unstable modes at
the smallest scales of the simulation, while at the same time the large
scales are kept unaffected by diffusivity. We have checked, by varying the value of the
artificial diffusion coefficient, that hyperdiffusion does not have any effect
on the turbulent diffusion coefficients that we measure. The use of
hyperdiffusivity is discussed further in \App{ch:hyperdiffusivity}. There the
hyperversions that we adopt for viscosity, mass diffusion and resistivity are
also presented.

In this section we also develop a method for being able to
treat numerically the motion of very tiny dust grains with friction times much shorter than
the computational time-step of the gas. This so-called short friction time
approximation is presented and discussed in the last part of the section.

\subsection{Gas Dynamics}

We consider the motion of gas and dust in the shearing sheet approximation
\citep[e.g.][]{GoldreichTremaine1978,Brandenburg+etal1995}. Here a local
coordinate frame corotating with the disc at a distance $r_0$ from the central
source of gravity is considered. The coordinate axes are defined as following.
The $x$-axis points always away from the central gravity source, and the $y$-axis
points in the direction of the Keplerian flow (as seen from a non-comoving
frame). The $z$-axis points perpendicular to the disc along the direction of
the angular velocity vector $\vc{\varOmega_0}$ of the orbital motion. In this
frame, the Keplerian flow velocity field has the linearized form $\vc{u}_0 =
-\frac{3}{2} \varOmega_0 x \hat{\vc{y}} \equiv u_y^{(0)} \hat{\vc{y}}$. We
choose to measure velocities relative to the Keplerian flow, $\vc{u} - \vc{u}_0
\rightarrow \vc{u}$. Such a transformation introduces new shear terms in the
equation of motion, but it has the advantage that the Keplerian velocity is zero
everywhere. The equation of motion relative to the main shear flow in the
shearing sheet approximation is
\begin{equation}
  \frac{\dpa \vc{u}}{\dpa t} = -(\vc{u}\cdot\nab)\vc{u}-
  u_y^{(0)} \frac{\dpa\vc{u}}{\dpa y} + \vc{f}(\vc{u}) -
  \frac{1}{\rho} \nab P + \frac{1}{\rho} \vc{J} \times \vc{B} +
  \vc{f}_{\rm \nu} (\vc{u},\rho)\, .
  \label{eq:gasmomentumeq}
\end{equation}
The first term on the right hand side of \Eq{eq:gasmomentumeq} is the advection
due to any velocity relative to the main shear flow, while the second term
covers the advection due to the shear flow. The function $\vc{f}(\vc{u})$ is
defined as
\begin{equation}
  \vc{f}(\vc{u}) = \pmatrix{2 \varOmega_0 u_y \cr -
  \frac{1}{2} \varOmega_0 u_x \cr 0}
\end{equation}
and is an effect of Coriolis force. The last three terms in
\Eq{eq:gasmomentumeq} are the pressure gradient force, the magnetic Lorentz
force (where the volume current density $\vc{J}$ is defined through Amp\`ere's law $\nab \times \vc{B} = \mu_0 \vc{J}$), and a hyperviscosity term based on the function $\vc{f}_{\rm
\nu}$ defined in \Eq{eq:fnu_3rd}. We ignore the effect of vertical gravity on
the gas, because we are interested in the ideal case to measure the
isotropy/non-isotropy of magnetorotational turbulence and the local transport
properties of the gas without introducing additional isotropy-breaking
effects. Ignoring the stratification effectively means that we are considering
the disc close to the mid-plane where the vertical gravity is vanishing. Future
work on the properties of dust diffusion in local shearing box simulations
should take the vertical stratification of the disc into account.

The evolution of the gas density $\rho$ is determined by the continuity
equation
\begin{equation}
  \frac{\dpa \rho}{\dpa t} = -u_y^{(0)} \frac{\dpa \rho}{\dpa y} -
  \rho \nab \cdot \vc{u} - \vc{u} \cdot \nab \rho + f_{\rm D}(\rho) \, ,
  \label{eq:gasconteq}
\end{equation}
where the first term on the right hand side is again an effect of advection due
to the main shear flow. The two next terms come from the standard term $\nab
\cdot (\rho \vc{u})$ from the continuity equation. In the last term we
include artificial mass diffusion through the function $f_{\rm D}$ defined in
\Eq{eq:fD_3rd}. An isothermal equation of state is used where the pressure
depends on the density as $P = c_{\rm s}^2 \rho$. Here $c_{\rm s}$ is the
constant sound speed.

The induction equation determines the evolution of the magnetic vector
potential $\vc{A}$. Evolving the vector potential has the advantage over
evolving the magnetic field $\vc{B} = \nab \times \vc{A}$ in that it maintains
a solenoidal magnetic field (i.e.\ $\nab \cdot \vc{B}=0$) at all times. The
induction equation in the shearing sheet approximation is
\begin{equation}
  \frac{\dpa \vc{A}}{\dpa t} = -u_y^{(0)} \frac{\dpa\vc{A}}{\dpa y} +
  \frac{3}{2} \varOmega_0 A_y \hat{\vc{x}} + \vc{u} \times \vc{B} +
  \vc{f}_\eta(\vc{A}) \, .
  \label{eq:inductioneq}
\end{equation}
The first term on the right hand side of \Eq{eq:inductioneq} is the advection
due to the main shear flow, while the second is the so-called magnetic
stretching term, another effect of shear \citep{Brandenburg+etal1995}. The two
last terms are the standard electromotive force and a resistivity term based on
the function $\vc{f}_\eta$ defined in \Eq{eq:feta_3rd}.

\subsection{Dust Dynamics}

The dust grains are considered as a fluid without any pressure support. Any
pressure gradient force on the dust due to collisions between dust grains and
gas molecules can also be ignored since the solid density of dust grains is so
large that the resulting acceleration is negligibly small.

In the fluid approach, the equation of motion for the dust velocity relative to
the Keplerian flow is
\begin{equation}
  \frac{\dpa \vc{w}}{\dpa t} = -(\vc{w}\cdot\nab)\vc{w} -
      u_y^{(0)} \frac{\dpa\vc{w}}{\dpa y} + \vc{f}(\vc{w}) +
      \vc{f}_\nu(\vc{w},n) - \frac{1}{\tau_{\rm f}} (\vc{w}-\vc{u}) +
      \vc{g}(x,y,z) \, .
  \label{eq:dustmomentumeq}
\end{equation}
The first four terms on the right hand side appear similar here as in the gas
momentum equation. The last two terms in
\Eq{eq:dustmomentumeq} are the drag force and an externally imposed force
field $\vc{g}$ that we use to drive a non-zero diffusion equilibrium in the dust
density. This is explained in more detail in \Sec{ch:diffusion_coefficient}.

We let the dust and the gas couple through a drag force proportional to, but in
the opposite direction of, the velocity difference between the dust and the
gas. The strength of the drag force is characterized by the friction time
$\tau_{\rm f}$. Any relative motion between dust and gas is damped by the
drag force with an
e-folding time of $\tau_{\rm f}$. The physics of the
dust grain and the gas enters in the expression of the friction time. When the
mean free path of the gas molecules is larger than the dust grain radius, the
dust grain is in the Epstein regime \citep[e.g.][]{Weidenschilling1977}. Here
the friction time of a spherical dust grain with radius $a_\bullet$ and solid
density $\rho_\bullet$ can be expressed as
\begin{equation}
  \tau_{\rm f} = \frac{a_\bullet \rho_\bullet}{c_{\rm s} \rho} \, ,
  \label{eq:tauf_epst}
\end{equation}
where $c_{\rm s}$ is the sound speed in the surrounding gas and $\rho$ is the
gas density. The sound speed and the gas density are approximately constant in
the unstratified and isothermal case, so we can treat $\tau_{\rm f}$ as
constant that depends only on the given particle radius and solid density.

Treating dust as a fluid is justified as long as the mean free path of the
fluid constituents is smaller than the typical dimensions of the system. In
the case of the gas, one compares the mean free path of the molecules with the
thickness of the disc $H_0$. For the dust grains the collisions among the
grains themselves are unimportant for determining a mean free path. Here the
collisions with the gas molecules are the important effect. The mean free path
for the dust grains can be defined as the distance one grain has to float with
respect to the gas before it has lost a significant fraction of its momentum.
For a spherical grain moving with a speed $w$ relative to the gas, this value
can be determined as $\ell = w \tau_{\rm f}$. The condition for treating dust
as a fluid is then that $\ell \ll H_0$. Because all motions are subsonic, we
can replace $w$ by $c_{\rm s}$ as an upper limit and get the expression
$\varOmega_0 \tau_{\rm f} \ll 1$ for the validity of the fluid approach.

In a typical solar nebula type protoplanetary disc, the scale height is of the
order of $H_0 \sim 10^{12}$ cm at $r_0=5\,{\rm AU}$, while the gas density can
be taken to $\rho_0 \sim 10^{-10}\,{\rm g\,cm^{-3}}$ at the same radial
distance. Then the connection between grain radius and dimensionless friction
time is
\begin{equation}
  a_\bullet = \varOmega_0 \tau_{\rm f} H_0 \frac{\rho_0}{\rho_\bullet} \sim 10^2
  \varOmega_0 \tau_{\rm f} \, {\rm cm} \, .
\end{equation}
This means that, as a rule of thumb, the value of the dimensionless friction time
corresponds to the radius of the dust grain measured in meters.

To preserve momentum the gas should be affected by a drag force $\vc{f}_{\rm
drag} = -\tau_{\rm f}^{-1} \rho_{\rm d}/\rho (\vc{u}-\vc{w})$ from the dust.
Here $\rho_{\rm d}/\rho$ is the dust-to-gas ratio. We shall ignore the
back-reaction drag force from the dust on the gas, because the dust-to-gas
ratio is small in the early stages of a protoplanetary disc.

We represent dust mass density by the number density $n$ of dust grains. The
continuity equation for the dust number density $n$ is
\begin{equation}
  \frac{\dpa n}{\dpa t} = -u_y^{(0)} \frac{\dpa n}{\dpa y} -
  n \nab \cdot \vc{w} - \vc{w} \cdot \nab n + f_{\rm D}(n) \, ,
  \label{eq:dustconteq}
\end{equation}
where we use artificial diffusion, through the function $f_{\rm D}(n)$ defined
in \Eq{eq:fD_3rd}, only to stabilize the numerical
scheme. By varying the value of the artificial diffusion coefficient $D_3$,
which is defined in \App{ch:hyperdiffusivity}, we have made sure that adding
artificial diffusion has no effect on the measured turbulent diffusion
coefficients. The value of $D_3$ needed to stabilize the runs are for all runs
several orders of magnitude below the measured turbulent diffusion coefficient.

We now have two possibilities to solve the dust equation of motion (eq.\
[\ref{eq:dustmomentumeq}]): For large particles, with friction times larger than
the Courant time-step ($\varOmega_0 \tau_{\rm f} > 0.001$, see
\Tab{t:rundescrps}), we can use the explicit integration scheme from the Pencil
Code. But for the smaller particle cases ($\varOmega_0 \tau_{\rm f} \ll 0.001$),
where the friction time is much shorter than the Courant time-step, we will use
the short friction time approximation, a semianalytical time integration that is
presented below.

\subsection{Short Friction Time Approximation}
\label{ch:shorttaufappr}

The radii of dust grains observed in protoplanetary discs are often on the
order of micrometers or even nanometers. The friction time of microscopic dust
grains in a protoplanetary disc is very short compared to the orbital period,
around a few minutes for the location of Jupiter in a typical solar nebula.
That is of course not a problem for nature, but the smallest scales of computer
simulations are many orders of magnitude larger than in nature, and thus the
computational time-step for an explicit code such as the Pencil Code is much
larger than the friction time of tiny dust grains. This causes a potential
problem in resolving both timescales at the same time. To follow the motion of
the tiniest dust grains, applying the explicit integration scheme as used in the
Pencil Code, a time-step must be chosen that is at least an order of magnitude
shorter than the friction time. Thus the computation time for simultaneously
following the evolution of gas and dust becomes prohibitively long. One can now
either use an implicit integration scheme, which would introduce further
problems, and also make major changes in the code necessary, or one can
use a kind of semianalytical integration scheme that works as follows.

For very short friction times, the dust is able to settle to an equilibrium
velocity, where the drag force is exactly balanced by the other
force terms that are present in the dust equation of motion (eq.\
[\ref{eq:dustmomentumeq}]), on a timescale that is
much shorter than the computational time-step of the gas. Thus it is possible,
under a few reasonable assumptions, to solve algebraically for the terminal
dust velocity as a function of gas velocity and density. To do this, we first
subtract the gas equation of motion (eq.\ [\ref{eq:gasmomentumeq}]) from the
dust equation of motion (eq.\ [\ref{eq:dustmomentumeq}]). This results in an
equation for the evolution of relative velocity $\vc{w}-\vc{u}$,
\begin{equation}
  \frac{\dpa(\vc{w}-\vc{u})}{\dpa t} + (\vc{w}\cdot\nab)\vc{w} - 
      (\vc{u}\cdot\nab)\vc{u} + 
      u_y^{(0)}\frac{\dpa(\vc{w}-\vc{u})}{\dpa y} = \vc{f}(\vc{w}-\vc{u})
      -\frac{1}{\tau_{\rm f}}(\vc{w}-\vc{u}) + \vc{g} +
      \frac{1}{\rho}(\nab P - \vc{J} \times \vc{B}) \, ,
  \label{eq:dwminudt}
\end{equation}
where the viscosity terms have been ignored since any real physical viscosity
is expected to be orders of magnitude weaker than the other force terms. We now
assume that the computational time-step of the gas is much longer than the
friction time, $\delta t \gg \tau_{\rm f}$. Here $\delta t$ will be given by
the Courant criterion. This criterion determines the maximum time-step that can
be taken by an explicit numerical scheme without becoming unstable. The allowed
time-step gets shorter with increasing grid resolution.
With the condition $\delta t \gg \tau_{\rm f}$, all terms from the gas equation
of motion can be considered to be constant for the duration of the acceleration
of the dust grain to its terminal velocity. This specifically also applies to the pressure
gradient force and the Lorentz force that are present also in
\Eq{eq:dwminudt}. Then we can search for a time-independent equilibrium
solution for $\vc{w}-\vc{u}$. We
expect any time-independent solution of \Eq{eq:dwminudt} to have a dust
velocity that is very close
to the gas velocity, because the short friction time couples the dust velocity
strongly to the gas velocity. Setting therefore $\vc{w}=\vc{u}$ in all other terms
than the drag force term (this is legitimized below) leaves the algebraic
equilibrium equation
\begin{equation}
  0 = -\frac{1}{\tau_{\rm f}}(\vc{w}-\vc{u}) + \vc{g} +
      \frac{1}{\rho}(\nab P - \vc{J} \times \vc{B}) \, .
  \label{eq:dwminudt_equi}
\end{equation}
Solving for $\vc{w}$ then yields
\begin{equation}
  \vc{w} = \vc{u} + \tau_{\rm f} \left[\vc{g} +
      \frac{1}{\rho}(\nab P - \vc{J} \times \vc{B}) \right] \, .
  \label{eq:w_sfta}
\end{equation}
Reinserting this solution into \Eq{eq:dwminudt} shows that it was reasonable to
ignore all advection, shear and Coriolis terms, while keeping the friction,
gravity, pressure gradient and Lorentz terms, as long as the friction time is
sufficiently short.

This is the short friction time approximation. The specific form of the short
friction time dust velocity approximation depends on the forces that are
assumed to work on the gas and on the dust, so \Eq{eq:w_sfta} is only valid for
the specific choice of force terms that are considered in this work. The
presence of gravity in the dust velocity approximation comes from only
considering gravity to work on the dust. This is is good for the purpose of
measuring the turbulent diffusion coefficient, whereas in nature gravity of
course affects both dust and gas. The gravity term would then drop out of the
short friction time approximation, but it would reappear in the form of the
vertical pressure gradient of the stratified gas. One must also take into
consideration that the dust velocity in \Eq{eq:w_sfta} is expressed as a
function of the resolved part of the gas velocity only. All unresolved small
scales would also contribute to the random motion of the tiny dust grains (as
would Brownian motion), but the important scales for turbulent transport are
the largest scales in the box, since they contribute most to the total gas
velocity field.

The equilibrium dust velocity given by the short friction time approximation
ensures that the relative velocity between dust and gas does not change on
timescales shorter than the computational time-step. That means that if the gas
is being accelerated, then the same amount of acceleration must be working on
the dust, and so the relative velocity between the dust and the gas stays
constant until sufficient time has passed for the pressure gradient force and
the Lorentz force to change at the computational timescale.

The $i$th component of the Lorentz force appearing in \Eq{eq:w_sfta} can be
rewritten as
\begin{equation}
  (\vc{J}\times\vc{B})_i = 
      \nabla_j \left( \frac{B_i B_j}{\mu_0} - \frac{B^2}{2 \mu_0} \delta_{ij}
      \right) \, ,
\end{equation}
where the first term in the parenthesis on the right hand side is due to
magnetic pressure and the second to magnetic tension. This allows the $i$th
component of the short friction time approximation dust velocity to be rewritten as
\begin{equation}
  w_i = u_i + \tau_{\rm f} \left[g_i +
      \frac{1}{\rho}\nabla_j \left(
      P \delta_{ij} + \frac{B^2}{2 \mu_0} \delta_{ij} - \frac{B_i B_j}{\mu_0} 
      \right) \right] \, .
  \label{eq:w_sfta_2}
\end{equation}
Thus dust grains move relative to the gas not only because of (additional)
gravity and (missing) pressure gradient force, but also due to (missing)
magnetic pressure gradient force and (missing) magnetic tension. We shall still
refer to the effect as pressure gradient trapping, even though the magnetic
tension term in \Eq{eq:w_sfta_2} does not mimic a pressure gradient.

We must stress again that the short friction time approximation is only valid
for small particles. If one considers larger bodies (e.g.\ $> 1$ mm at
$r_0=5\,{\rm AU}$ in a typical solar nebula), first the Coriolis forces and
then the advective transport terms can no longer be ignored. For these
particles we directly integrate the dust equation of motion (eq.
[\ref{eq:dustmomentumeq}]) together with the other dynamical equations. With
even larger objects finally the fluid approach fails as soon as $\ell
> H_0$. In this case one has to apply a particle algorithm to follow the dust
evolution \citep[e.g.][]{KlahrHenning1997}.

\section{DIFFUSION COEFFICIENT}
\label{ch:diffusion_coefficient}

In this section we describe how we calculate the diffusion coefficient of dust
grains embedded in a turbulent gas. We do this by comparing the results of numerical
simulations with analytical solutions to the non-turbulent flow equations that
include a parametrized diffusion.

If the turbulent motion of the gas and the dust has not been resolved, the
continuity equation of the dust would have to incorporate an explicit diffusion
term,
\begin{equation}
  \frac{\dpa n}{\dpa t} = 
      -\nab \cdot \left[ (\vc{w}+u_y^{(0)} \hat{\vc{y}}) n - D_{\rm t} \rho
      \nab \left( \frac{n}{\rho} \right) \right] \, .
  \label{eq:conteq_dust_lam}
\end{equation}
The gas flow is here assumed to be completely stationary, and the only effect
of the non-resolved turbulence is through the parametrized diffusion term. The
continuity equation is written in a conservative form where the diffusion flux
is proportional to and in the opposite direction of the gradient of the
dust-to-gas ratio. This is the way turbulent diffusion is normally assumed to
act \citep[see e.g.][]{Dubrulle+etal1995}.

The task now is to find a way to extract $D_{\rm t}$ from the non-stationary
turbulent motion found in computer simulations. This is only possible if $\nab
n$ is not zero everywhere, as otherwise the diffusion coefficient does not
enter \Eq{eq:conteq_dust_lam} at all for a constant $\rho$. One can now either
follow the time dependent diffusion of an initial dust concentration somewhere
in the center of the box or look for a time independent equilibrium solution.
The first approach has the disadvantage that it is difficult to obtain good
statistics, as one has always a very special distribution with a distinct
wavelength, whereas the turbulence could act on all scales. Therefore we use
the latter possibility and search for an equilibrium solution where we can
achieve much better statistics.

We force an equilibrium solution with a non-zero dust density gradient by exposing the
grains to an external force field $\vc{g}$. Depending on its specific form,
this force field will eventually result in an equilibrium where the pile-up of
dust grains imposed by $\vc{g}$ is balanced completely by mass diffusion in the
opposite direction. By comparing the
analytical expression for the equilibrium dust number density, whose only free
parameter is $D_{\rm t}$, to the equilibrium density obtained when the
turbulence is resolved in computer simulations, we can derive the turbulent
diffusion coefficient. We will often refer to the external force field simply
as gravity because of the qualitative similarities to real gravity.

First the diffusion in the $z$-direction is considered. Here we define a
vertical gravity field
\begin{equation}
  g_z = -g_0 \sin(k_z z) \, ,
  \label{eq:gz}
\end{equation}
where $k_z = 2\pi/L_z$ in
order to have periodic boundaries in the vertical direction. Here $L_z$ is
the vertical extent of the box, and $z$ is defined to lie in the interval between
$-\frac{1}{2} L_z$ and $\frac{1}{2} L_z$. Using periodic
boundary conditions demands that we use a periodic force field in order to have
a periodic equilibrium solution. The gravity
field defined in \Eq{eq:gz} is linear around the mid-plane, as the gravity
field normally considered for thin discs also is, but away from the mid-plane
it becomes zero again on the top and bottom boundaries of the box. Such a force
gives a periodic dust distribution to determine the turbulent viscosity
coefficient from (we will show below that the equilibrium logarithmic
dust density becomes cosinusoidal with $z$). For a normal thin disc vertical
gravity field, $g_z=-\varOmega_0^2 z$, the equilibrium logarithmic dust
density becomes quadratic with $z$, which is obviously not periodic.

To find the equilibrium dust number density, we solve now 
\Eqs{eq:dustmomentumeq}{eq:conteq_dust_lam} for
$\dpa \vc{w}/\dpa t=\dpa n/\dpa t=\vc{u}=w_x=w_y=0$, $w_z = w_z(z)$ and $n=n(z)$.
This yields the differential equation system
\begin{eqnarray}
  0 &=& 
  -w_z\frac{\dpa w_z}{\dpa z} - \frac{1}{\tau_{\rm f}} w_z -g_0 \sin(k_z z) \, ,
  \label{eq:dustmomentumeq_z}
  \\
  0 &=& -\frac{\dpa}{\dpa z} \left( w_z n - D_z^{\rm (t)} \frac{\dpa n}{\dpa z}
  \right) \, ,
  \label{eq:dustconteq_z}
\end{eqnarray}
where we neglect the $\rho$-dependence in the diffusion term, because the
turbulent
gas density fluctuations are very small, as expected in subsonic turbulence.
For any sufficiently short friction time, the advection term in
\Eq{eq:dustmomentumeq_z} can be safely ignored, leaving only the algebraic
equation
\begin{equation}
  0 = - \frac{1}{\tau_{\rm f}} w_z -g_0 \sin(k_z z)
\end{equation}
with the solution
\begin{equation}
  w_z = -\tau_{\rm f} g_0 \sin(k_z z) \, .
  \label{eq:dustvelocity_eq_z}
\end{equation}
Inserting \Eq{eq:dustvelocity_eq_z} into \Eq{eq:dustmomentumeq_z} shows that
the advection term is fully negligible for $\tau_{\rm f}^2 g_0 k_z \ll 1$.

The equilibrium solution to the continuity equation must now be able to
continuously replace material that is being transported towards the mid-plane
by new material transported away from the mid-plane by diffusion. It is seen
that \Eq{eq:dustconteq_z} has the general solution
\begin{equation}
  \ln n = \frac{1}{D_z^{\rm (t)}} \int w_z(z) \de z
  \label{eq:lnnequi_z}
\end{equation}
for any integrable function $w_z(z)$. Here we have assumed that there is no net
flux of dust grains $\overline{w_z}=0$ by setting the contents of the
parenthesis on the right hand side of \Eq{eq:dustconteq_z} equal to zero.
Inserting the equilibrium dust velocity from \Eq{eq:dustvelocity_eq_z} into the
integral in \Eq{eq:lnnequi_z} gives the equilibrium logarithmic dust number
density as
\begin{equation}
  \ln n = \ln n_1 + \frac{\tau_{\rm f} g_0}{k_z D_z^{\rm (t)}} \cos(k_z z) \, ,
  \label{eq:lnnz_equi}
\end{equation}
where $\ln n_1$ is an integration constant that corresponds physically to the
logarithmic number density at $z=\pm \frac{1}{4} L_z$. The amplitude of the cosine distribution depends
only on friction time, gravity strength and gravity wave number, which are all
known input parameters, and the unknown value of the turbulent diffusion
coefficient in the vertical direction. Thus the value of the turbulent
diffusion coefficient can be determined uniquely from this amplitude.

The number density distribution in \Eq{eq:lnnz_equi} is not normalized. The
connection between $n_1$ and the column density
$\varSigma_0$ is
\begin{equation}
  \varSigma_0 = \int_{-1/2 \, L_z}^{1/2 \, L_z} n_1 
      \exp\left[ \frac{\tau_{\rm f} g_0}{k_z D_z^{\rm (t)}} \cos(k_z z)\right]
      \de z = \frac{2 \pi n_1}{k_z} \frac{1}{\pi} \int_0^{\pi}
      \exp\left[ \frac{\tau_{\rm f} g_0}{k_z D_z^{\rm (t)}} \cos(k_z z)\right]
      \de (k_z z) \, ,
      \label{eq:Sigma0n1int}
\end{equation}
where the last equality holds because the cosine function is symmetric in $z$.
The modified Bessel function of the first kind of order $m$ is defined as
\begin{equation}
  I_m(x) = 
      \frac{1}{\pi} \int_0^\pi \ee^{x \cos \theta} \cos(m \theta) \de \theta
  \, ,
  \label{eq:Bess}
\end{equation}
so the connection between $\varSigma_0$ and $n_1$ becomes simply
\begin{equation}
  \varSigma_0 = \frac{2 \pi n_1}{k_z}
      I_0\left(\frac{\tau_{\rm f} g_0}{k_z D_z^{\rm (t)}}\right) \, .
\end{equation}
Isolating $n_1$ finally yields
\begin{equation}
  n_1 = \frac{k_z \varSigma_0}{2 \pi 
      I_0\left(\frac{\tau_{\rm f} g_0}{k_z D_z^{\rm (t)}}\right)} \, .
  \label{eq:n1_S0}
\end{equation}
For infinite diffusion $D_z^{\rm (t)} \rightarrow \infty$, the argument of the
Bessel function in \Eq{eq:n1_S0} is zero, and using $I_0(0) = 1$ from
\Eq{eq:Bess}, we get $\varSigma_0 = n_1 L_z$. Thus $n_1=n_0$, where
$n_0$ is the average dust number density in the box, as expected for the
special case of infinite diffusion. In the case of a finite diffusion
coefficient, $n_1 \ne
n_0$.

For the radial $x$-direction, a similar sinusoidal gravity field can be defined
to give the equilibrium dust density as
\begin{equation}
  \ln n = \ln n_1 + \frac{\tau_{\rm f} g_0}{k_x D_x^{\rm (t)}} \cos(k_x x) \, ,
  \label{eq:lnnx_equi}
\end{equation}
formally identical to the vertical case. The derivations are given in
\App{ch:radial_diffusion_eq}. With \Eqs{eq:lnnz_equi}{eq:lnnx_equi} we are
armed with two powerful analytical expressions for the number density
distribution of dust grains in diffusion equilibrium with an externally imposed
force field. By comparing computer simulations of magnetorotational turbulence
with these analytical results, we can extract the turbulent diffusion
coefficient of the dust grains in both the vertical and the radial directions
independently. The next sections describe the setup of the simulations and the
results that we get.

\section{UNITS AND BOUNDARY CONDITIONS}
\label{ch:units}

We adopt non-dimensional variables by measuring velocities relative to the
isothermal sound speed, $[\vc{u}]=[\vc{w}]=c_{\rm s}$, and densities relative
to the initial density in the box, $[\rho]=\rho_0$; $[n]=n_0$. The unit of
dust-to-gas ratio $\epsilon_{\rm d}$ is $[\epsilon_{\rm d}]=\epsilon_0=m_0
n_0/\rho_0$, where $m_0$ is the mass of the individual dust grains. Time is
measured in units of inverse Keplerian angular speed, $[t]=\varOmega_0^{-1}$,
although often stated in orbits $T=2\pi\varOmega_0^{-1}$. The unit of magnetic
field is $[\vc{B}]=c_{\rm s} \sqrt{\mu_0 \rho_0}$. Derived from these basic
units are the unit of distance $[\vc{x}]=c_{\rm s} \varOmega_0^{-1}$ and the
unit of magnetic vector potential $[\vc{A}]=c_{\rm s}^2 \varOmega_0^{-1}
\sqrt{\mu_0 \rho_0}$. The unit of turbulent viscosity and turbulent diffusion
coefficient can also be derived from the basic units to be $[\nu_{\rm
t}]=[D_{\rm t}]=c_{\rm s}^2 \varOmega_0^{-1}$. In these units the turbulent
viscosity and the turbulent $\alpha$-value take the same numerical value.

We choose a box length of $2 \pi$ in all directions. In order to keep the
background shear flow subsonic at all points we choose the arbitrary
normalization $\varOmega_0=0.2$. We have checked by setting $\varOmega_0$ to
unity that the evolution of the simulations indeed scale with the value of
$\varOmega_0$, and thus that the scale-free diffusion coefficients and
$\alpha$-values are independent of the choice of $\varOmega_0$.

Periodic boundary conditions are applied in all directions. Connected points at
the periodic $x$-boundary have a time-dependent shift as is appropriate in the
shearing sheet approximation.

\section{EVOLUTION OF GAS}
\label{ch:evolution_gas}

As an initial condition, we perturb the gas velocity components with random
fluctuations of amplitude $\delta\vc{u} \sim 10^{-3}$. The toroidal component
of the magnetic vector potential is perturbed by a standing cosine wave $A_y =
A_0 \cos(k_x x) \cos(k_y y) \cos(k_z z)$ of amplitude $A_0=0.2$ and wave
numbers $k_x = k_y = k_z = 1$. The resulting vertical component of the
magnetic field is $B_z = -A_0 k_x \sin(k_x x) \cos(k_y y) \cos(k_z z) =
B_0(x,y) \cos(k_z z)$. Such a wave is unstable to shear if $k_z$ is
sufficiently small (i.e.\ at sufficiently large scales). As shown by
\cite{BalbusHawley1991}, the wave number interval for instability of the
vertical magnetic field component is $0 <k_z < \sqrt{3} \varOmega_0/v_{\rm A}$,
where the Alfv\'en speed is defined as $v_{\rm A}^2=B_0^2/(\mu_0 \rho_0)$. For
$0< |B_0| < 0.2$, the upper limit wave number is always larger than around
$k_z=\sqrt{3}$, so $k_z=1$, the largest scale present in the simulation, is
well within the unstable regime.

We run simulations in two different resolutions, $64^3$ and $128^3$. In
\Tab{t:rundescrps} the parameters that are used in the different runs are
listed. As there is no back-coupling from the dust on the gas, the gas
evolution depends mainly on resolution, since the high resolution runs require
less artificial diffusivity. The dust only affects the gas through its
contribution to the computational time-step.

\subsection{Self-Sustained Turbulence}

Initially the magnetic and kinetic energies in the box increase, but the
increase stops after around half an orbit, and then the magnetic and kinetic
energies fall slowly to an equilibrium state during a transition time of around
ten orbits. In the equilibrium state the turbulence is self-sustained, in the
sense that energy is pumped from the gravitational field primarily into
magnetic energy (via the magnetorotational instability). The Lorentz force
transfers some of this magnetic energy into turbulent kinetic energy which
again transfers energy back into the magnetic field in a dynamo process.
Finally the energy is dissipated through resistivity and viscosity. The whole
process is sketched in \cite{Brandenburg+etal1995}. Because we assume an
isothermal equation of state, there is no heating of the gas due to dissipative
processes.

\begin{figure}
  \includegraphics[width=17.3cm]{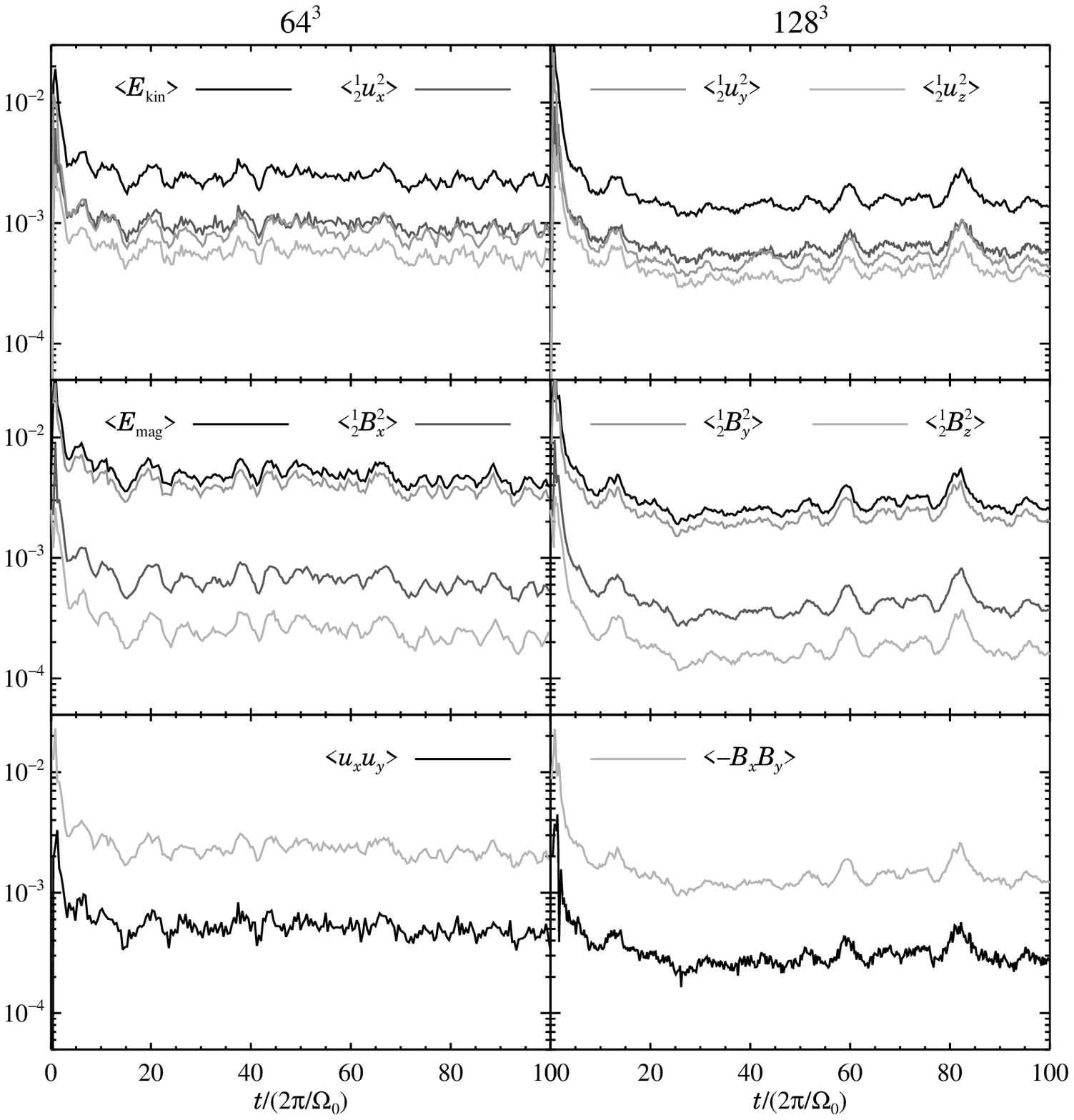}
  \caption{Evolution of various turbulence parameters for a $64^3$ run (left
  panels) and a $128^3$ run (right panels). The top panels show the evolution
  of total kinetic energy and its directional components. The radial and
  toroidal directions have comparable values of kinetic energy, whereas the
  vertical direction has around a factor of two less. The magnetic energy
  (middle panels) is completely dominated by the toroidal magnetic field. The
  $uy$ component of the Reynolds and Maxwell stresses (lower panels) is
  effectively a measure of the turbulent viscosity. The magnetic stresses
  are around four times higher than the kinetic stresses.}
  \label{fig:turbulence_pars}
\end{figure}

The time evolution of kinetic energy components, magnetic energy components, and
Reynolds and Maxwell stresses is shown in \Fig{fig:turbulence_pars} for a time
span of 100 orbits. All turbulence parameters are approximately constant in
time, within a certain fluctuation interval, and show no sign of decaying after
the steady state has set in after around ten orbits. Most of the kinetic energy
(top panels) is present in the horizontal components of the velocity field, which is
always measured relative to the Keplerian flow, whereas the vertical component
contains a factor of two lower kinetic energy \citep[this anisotropic trend is
normal to MRI simulations, see e.g.][]{Hawley+etal1995}. For the magnetic
energy (middle panels), almost the entire energy is kept in the toroidal
component of the magnetic field. The ratio between kinetic and magnetic
energies stays approximately constant in time with the magnetic energy being a
factor of around two higher than the kinetic energy. The Reynolds and Maxwell
stresses (shown in the two bottom panels) can be converted into a turbulent
viscosity and normalized to a turbulent $\alpha$-value of
\cite{ShakuraSunyaev1973}. These values are shown for the different runs in the
second and third columns of \Tab{t:nuturb}. The magnetic $\alpha$-value is
around a factor of four times the non-magnetic, so most angular momentum
transport happens because of magnetic fields. In the shearing sheet
approximation the Keplerian background velocity is linear in space, so there is
no pile-up of angular momentum anywhere in the box.

\begin{figure}
  \includegraphics[width=8.3cm]{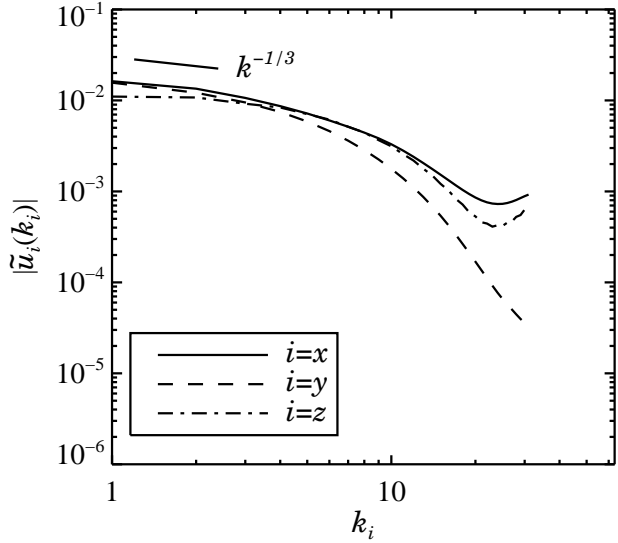}
  \caption{Fourier spectrum of the velocity components of the gas for a $64^3$
  resolution run, averaged from 10 to 100 orbits. A Kolmogorov $k^{-1/3}$ line
  is shown for reference. Both the radial and the toroidal components show a
  Kolmogorov-like behavior on large scales, whereas the vertical component is
  flatter. At small scales dissipation becomes important. The radial and
  vertical directions show a rise in power on the very smallest scales.}
  \label{fig:power64_uxyz}
\end{figure}

The magnetorotational instability injects energy at the largest scales of the
box. The smaller scales are then set in motion as the large scale motion
cascades down to smaller and smaller scales. Under the assumption that there is
no pile-up of kinetic energy at any scales, the Fourier spectrum should obey a
Kolmogorov-law $\tilde{u}(k) \sim k^{-1/3}$. The Fourier spectra of all velocity
components for $64^3$ and $128^3$ runs are shown in
\Figs{fig:power64_uxyz}{fig:power128_uxyz}. For reference a $k^{-1/3}$ line is
shown. The spectra are averages taken from 10 to 100 orbits. At large scales,
the power spectra approximately obey a Kolmogorov law, but at smaller scales,
where dissipation becomes important, the slope becomes steeper. There is some
excess power at the very smallest scales, especially for the $64^3$ run. This
is due to unstable modes at the smallest scales of the box. Curiously the
excess power is only present in the radial and vertical directions and not in
the toroidal direction, but this may be an effect of the shearing out of all
variables along $y$. The power in the small scale modes is still negligible
compared to the large scales, so the rise in power does not influence the
diffusion of the dust. These rises in power are typical for simulations with a
low diffusivity, see e.g.\ \cite{Haugen+etal2004}. According to mixing length
theory, the contribution from the different length scales to the total turbulent
diffusion coefficient scales as $D_k \sim \tilde{u}_k/k \sim k^{-4/3}$, so the
largest scales of the box are expected to give the dominating contribution to
the total turbulent diffusion coefficient.

One also sees from \Figs{fig:power64_uxyz}{fig:power128_uxyz} that in both cases the
vertical velocity amplitude on the large scales is smaller than the radial and
toroidal velocity amplitude at large scales. This gives already a hint that
vertical turbulent diffusion might be weaker than radial turbulent diffusion.

\section{EVOLUTION OF DUST}
\label{ch:evolution_dust}

\begin{figure}
  \begin{center}\includegraphics[width=12.0cm]{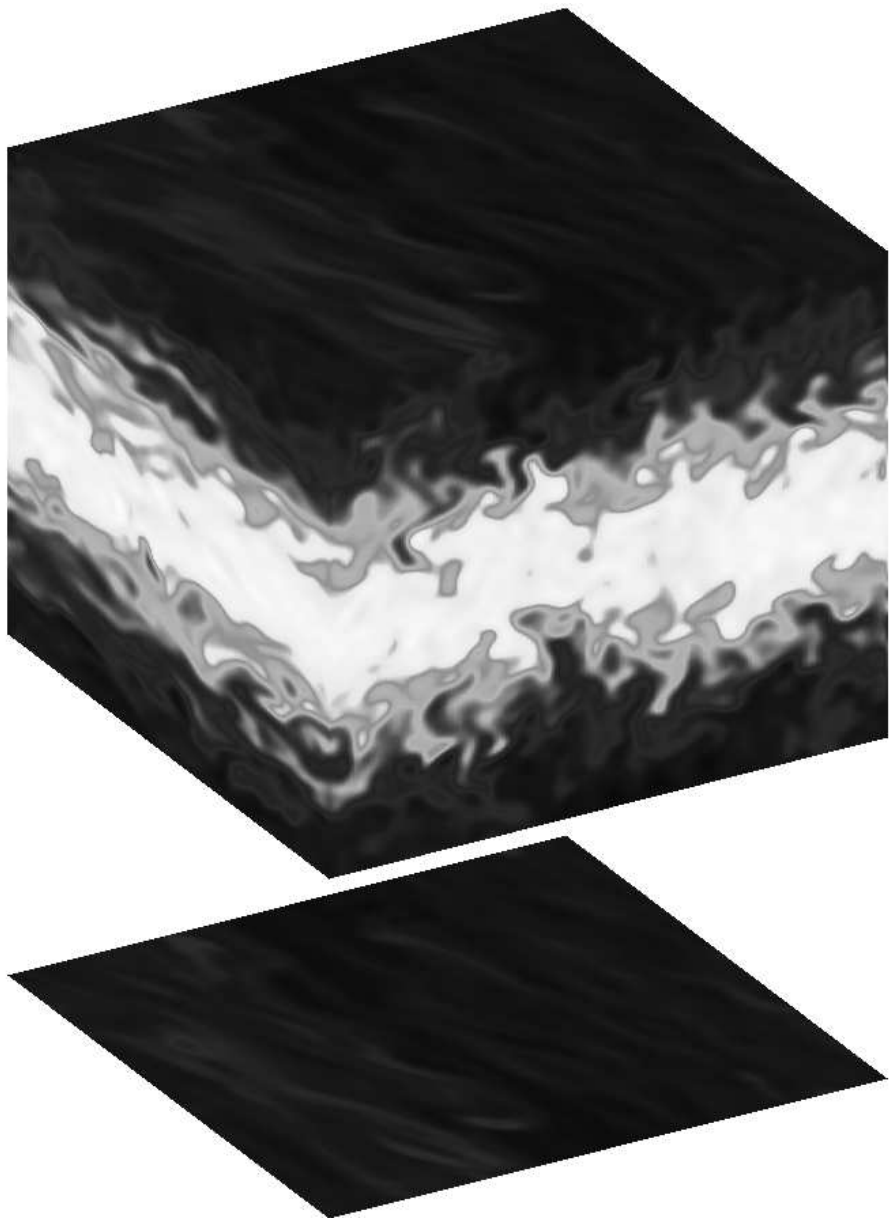}\end{center}
  \caption{Dust density contours at the sides of the simulation box for the
  short friction time run 128a\_z. The radial direction is towards the right
  while the shearing direction is towards left. The dust is concentrated
  around the mid-plane due to a vertical gravity acting only on the dust.
  Turbulent transport alone prevents the further vertical settling of the dust
  layer. This configuration is statistically unchanged for at least one
  hundred orbits.}
  \label{fig:dust_box}
\end{figure}

The dust is initially at rest and has a constant number density $n(x,y,z)=n_0$.
It is then set free to evolve under the influence of friction with the gas and the
imposed gravity field. The dust begins to concentrate near the center of
gravity (horizontal mid-plane, with $z=0$, for vertical gravity, vertical
mid-plane, with $x=0$, for radial gravity), but eventually an equilibrium
configuration is reached where the turbulent diffusion prevents further
concentration. This situation is shown in \Fig{fig:dust_box} for a $128^3$ run
with $\varOmega_0 \tau_{\rm f}=2 \times 10^{-7}$ and vertical gravity. The run
is labeled 128a\_z in \Tab{t:rundescrps}, and the friction time corresponds to
tiny dust grains or molecules with radii of 0.2 micrometers in a typical 
solar nebula. The plot shows dust density contours at the sides of the
simulation box. The turbulent motion is clearly visible, and the resulting
turbulent diffusion is the only reason why there is no further settling of the
dust layer towards the mid-plane. The amplitude of the concentration around the
mid-plane is maintained approximately constant for the entire duration of the
simulation (one hundred orbits).

\subsection{Diffusion Timescale}

Before proceeding with measuring diffusion coefficients, we will
first make an estimate of the
time it takes to get from a constant dust density to the
equilibrium where sedimentation is balanced by turbulent diffusion. We consider
the case of vertical gravity. The logarithmic dust density must rise from
an initial value $\ln n_0$ to the equilibrium value given by \Eq{eq:lnnz_equi}.
When the amplitude of the equilibrium cosine function is small, $A_{\ln n}\ll1$, then
we can assume that $n_1 \approx n_0$. The increase in logarithmic dust density
is then simply
\begin{equation}
  \Delta \ln n = \frac{\tau_{\rm f} g_0}{k_z D_z^{\rm (t)}} \cos (k_z z) \, .
  \label{eq:Deltalnn}
\end{equation}
This increase is caused by the vertical sedimentation. In the short friction
time approximation the dust velocity can be written as $w_z = -\tau_{\rm f}
g_0 \sin(k_z z)$. The change in logarithmic dust density due to vertical
settling can be approximated with the expression
\begin{equation}
  \frac{\dpa \ln n}{\dpa t} = -\frac{\dpa w_z}{\dpa z} 
      = k_z \tau_{\rm f} g_0 \cos(k_z z) \, .
  \label{eq:dlnndt_sedi}
\end{equation}
Here we have ignored the advection of mass for simplicity. The diffusion
timescale $t_{\rm D}$ can now be estimated by dividing \Eq{eq:Deltalnn} with
\Eq{eq:dlnndt_sedi}. This yields
\begin{equation}
  t_{\rm D} = \frac{1}{k_z^2 D_z^{\rm (t)}} \, .
  \label{eq:tD}
\end{equation}
Rewriting the diffusion coefficient in dimensionless units as $D_z^{\rm (t)} =
\delta_z^{\rm (t)} c_{\rm s}^2 \varOmega_0^{-1}$, the diffusion timescale can
be written as $\varOmega_0 t_{\rm D} = [(c_{\rm s} \varOmega_0^{-1} k_z)^2
\delta_z^{\rm (t)}]^{-1}$. With $\varOmega_0=0.2$, $k_z=1$ and $\delta_z^{\rm
(t)}=0.002$, the diffusion timescale is around three orbits. The diffusion
timescale for radial diffusion is completely equivalent to \Eq{eq:tD}.

For a linear gravity field, \cite{DullemondDominik2004} derive a diffusion
timescale similar to \Eq{eq:tD}. On the other hand, \cite{Dubrulle+etal1995}
state a diffusion timescale of $1/(\varOmega_0^2 \tau_{\rm f})$. This
expression is actually a gravitational settling timescale that determines the
amount of time it takes to increase the dust density in the mid-plane
significantly due to gravity. Since the diffusion equilibrium sets in at
very modest mid-plane overdensities for small dust grains, the timescale for
such grains to reach
diffusion equilibrium is much shorter than the gravitational settling timescale.

\begin{figure}
  \includegraphics[width=8.3cm]{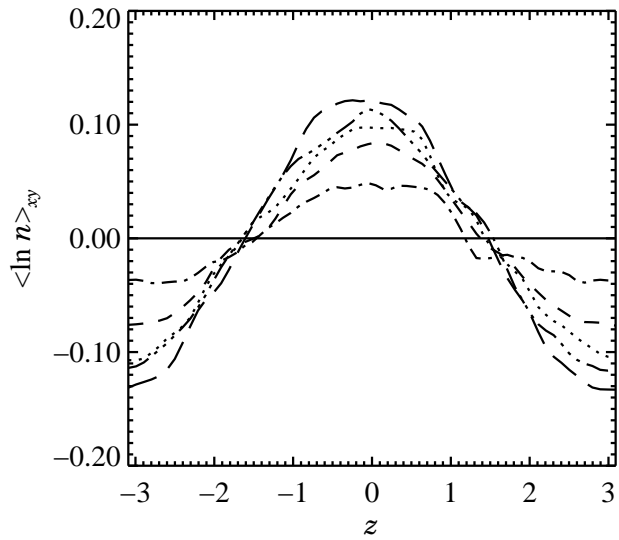}
  \caption{The logarithmic dust density of the run 64a\_z, averaged over $x$ and
  $y$, at different times. The curves are each separated by
  two orbits going from $t=0$ (full line) to $t=10$ orbits (long-dashed line).
  The approach to equilibrium happens on a timescale of a few orbits, in good
  agreement with the analytical estimate of the diffusion timescale that
  is presented in the text.}
  \label{fig:lnn_vs_zt}
\end{figure}

\begin{figure}
  \includegraphics[width=8.3cm]{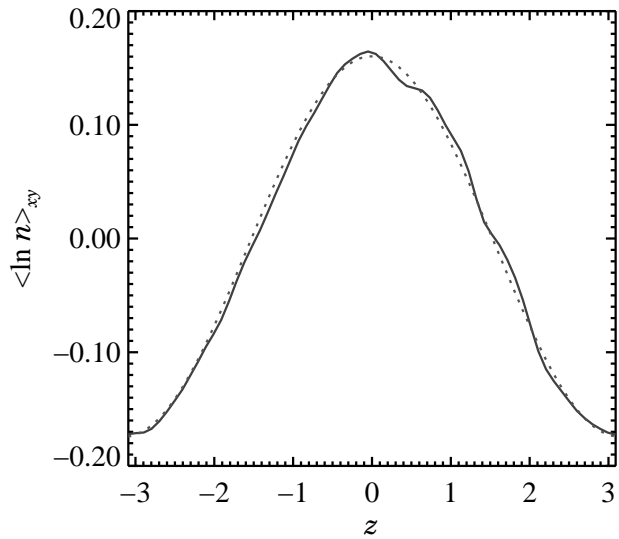}
  \caption{The logarithmic dust density averaged over $x$ and $y$ as function
  of vertical height $z$ (full line) and a cosine fit (dotted line). The cosine
  fit is in excellent agreement with the data. This shows that the turbulent
  transport is indeed well described as diffusion. Shown here is for the short
  friction time run 64a\_z at a time of 26 orbits. The fit quality (defined in
  the text) is $Q\approx0.005$.}
  \label{fig:lnn_vs_z_taus1e-6_z}
\end{figure}

In \Fig{fig:lnn_vs_zt} we plot the evolution of the logarithmic dust density
averaged over the $x$- and $y$-directions for the run 64a\_z. Starting at a
time of zero orbits, curves are shown at two orbits time separation up to a
time of ten orbits. The timescale to reach diffusion equilibrium is evidently
around a few orbits (the saturated state is shown in
\Fig{fig:lnn_vs_z_taus1e-6_z}). This is in good agreement with the analytical
estimates given above.

\subsection{Measured Turbulent Diffusion Coefficients}

\begin{figure}
  \includegraphics[width=8.3cm]{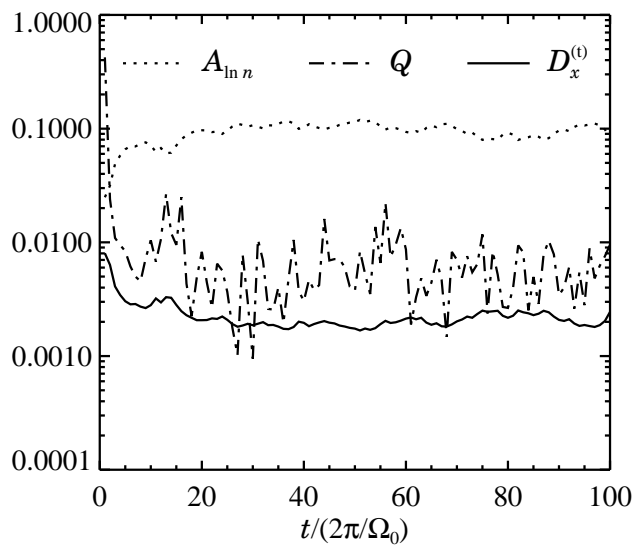}
  \caption{Time evolution of the fitted cosine amplitude (dotted line), the fit
  quality (dash-dotted line) and the derived radial turbulent diffusion
  coefficient (full line) for the short friction time run 64a\_x with radial
  gravity. Both the fit amplitude, and hence the turbulent diffusion
  coefficient,
  are approximately constant in time, although small variations are seen. The
  fit quality is generally excellent, but it fluctuates with around an order of
  magnitude during the 100 orbits shown here. Compare with
  \Fig{fig:ampl_Q_Dt_taus0.1_x} which shows the evolution of the same
  variables for an intermediate friction time run.}
  \label{fig:ampl_Q_Dt_taus1e-6_x}
\end{figure}

We now turn to measuring the turbulent diffusion coefficient from the
equilibrium configuration that is illustrated in \Fig{fig:dust_box}.
According to
\Eqs{eq:lnnz_equi}{eq:lnnx_equi}, the equilibrium logarithmic dust density
should be a cosine function if diffusion is the proper description of the
turbulent transport. As an example of how the vertical diffusion coefficient is
measured, we show in \Fig{fig:lnn_vs_z_taus1e-6_z} the logarithmic dust density
averaged over the radial and toroidal directions for the run 64a\_z at a time
of $t=38$ orbits. Also shown is the minimum $\chi^2$ cosine fit (dotted line).
The fit is excellent, and this shows that here the turbulent transport of the
dust grains is well-described as diffusion. In \Fig{fig:ampl_Q_Dt_taus1e-6_x}
we plot, for the same run, the full time evolution of the amplitude of the
best-fit cosine function and the quality of the fit, $Q$. The fit quality is
defined as
\begin{equation}
  Q \equiv \frac{\sum_i [\langle\ln n\rangle_{xy}(z_i) - (\ln n)_{\rm fit}(z_i)]^2}
      {\sum_i [\langle\ln n\rangle_{xy}(z_i)]^2} \, .
\end{equation}
Here the sum is taken over the entire vertical direction and
$\langle\ldots\rangle_{xy}$ is used
to denote the average taken over the $x$- and $y$-directions. In
\Fig{fig:ampl_Q_Dt_taus1e-6_x} we also plot the
turbulent diffusion coefficient derived from the amplitude using
\Eq{eq:lnnz_equi}. Both the amplitude and the turbulent diffusion coefficient
stay approximately constant in time. The fit quality fluctuates by more than an
order of magnitude, but is generally very good ($Q$ is less than 0.02 at all times). A
similar behavior is found for all runs with short friction times, both for
vertical and for radial gravity.

\begin{figure}
  \includegraphics[width=8.3cm]{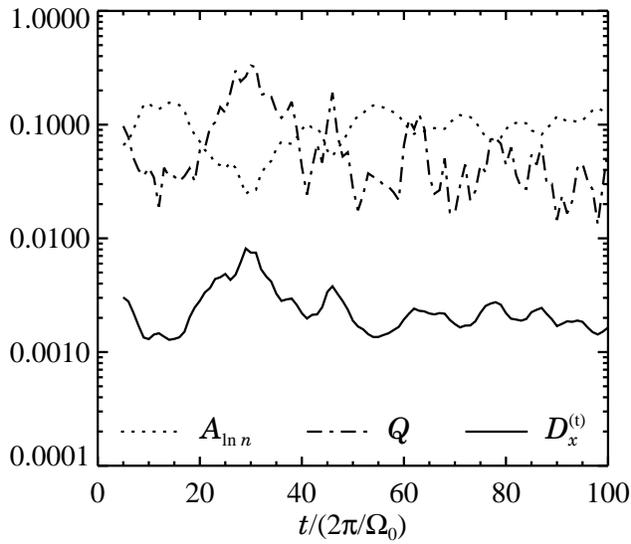}
  \caption{Same as \Fig{fig:ampl_Q_Dt_taus1e-6_x}, but for the intermediate
  friction time run 64c\_x. Obviously the cosine amplitude and the derived
  turbulent diffusion coefficient change much more violently with time. The fit
  quality is also a lot worse. The average diffusion coefficient is actually
  the same as for the short friction time run shown in
  \Fig{fig:ampl_Q_Dt_taus1e-6_x}, but the poor fit quality here means that the
  diffusion description of turbulent transport is not as good as it is in the
  short friction time runs.}
  \label{fig:ampl_Q_Dt_taus0.1_x}
\end{figure}

We have also run simulations without the short friction time approximation. Here
the friction time must be at least a few times longer than the computational
time-step of the gas in order to resolve the frictional acceleration in our
explicit numerical scheme, but shorter than an orbital period for the fluid
approach to be valid. We shall refer to such values of the friction time
as intermediate friction times. Simulations with a freely evolving dust
velocity serve both the purpose of showing in how far the short friction time
approximation is valid, and also how the turbulent diffusion coefficient behaves when
the friction time becomes larger and acceleration effects come into play.
Remember that the short friction time approximation assumes that the dust
grains can always reach an equilibrium velocity in one computational time-step.
Hence effects such as vortex trapping in turbulent eddies
are not possible in the short friction time approximation. The time evolution of
cosine amplitude, fit quality and turbulent diffusion coefficient for the
intermediate friction time run 64c\_x (with radial gravity and a friction time
of $\varOmega_0 \tau_{\rm f}=0.02$ corresponding to dust grains with radii of a
few centimeters) is shown in \Fig{fig:ampl_Q_Dt_taus0.1_x}. Here the amplitude
changes a lot with time, and the fit quality $Q$ rises above 0.1 on several
occasions. Apparently diffusion is not at all times a good description of the
turbulent transport in this run, even though the grains are still relatively
well-coupled to the gas. Nevertheless, the time averaged diffusion coefficient
in \Fig{fig:ampl_Q_Dt_taus1e-6_x} and \Fig{fig:ampl_Q_Dt_taus0.1_x} is
approximately the same as for the small grains.

\begin{figure}
  \includegraphics[width=\linewidth]{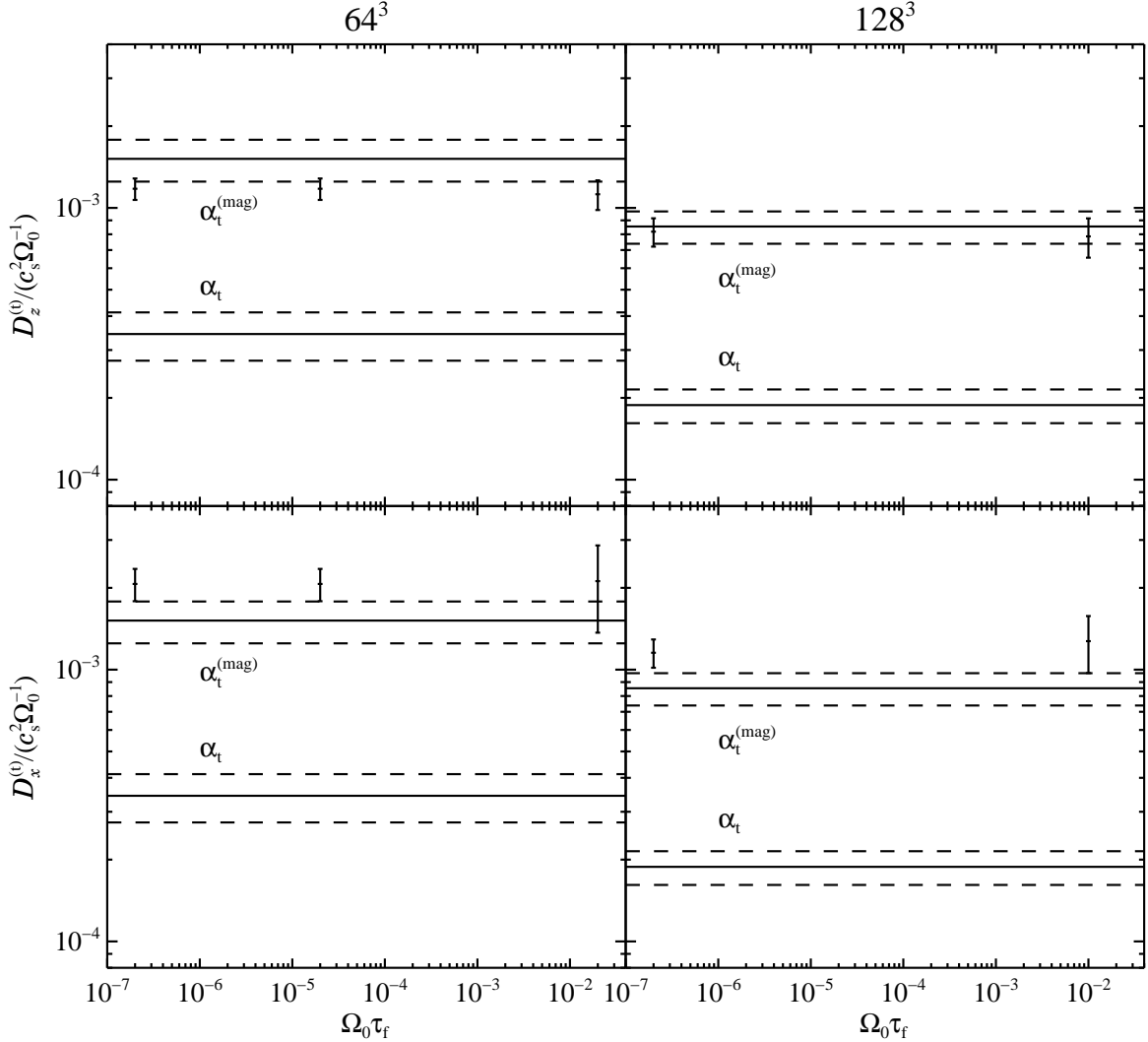}
  \caption{The measured turbulent diffusion coefficient as a function of
  $\varOmega_0 \tau_{\rm f}$ for a resolution of $64^3$ (left panels) and
  $128^3$ (right panels). The vertical diffusion coefficient is shown in the
  top panels, while the radial diffusion coefficient is shown in the bottom
  panels. For reference the turbulent $\alpha$-values based on both the
  Reynolds stress and the Maxwell stress are shown including their 1-$\sigma$
  fluctuation intervals. The radial diffusion coefficient is comparable to the
  sum of the turbulent $\alpha$-values and is around $70\%$ higher than the
  vertical diffusion coefficient.}
  \label{fig:Dt_vs_tauf_panel}
\end{figure}

The measured turbulent diffusion coefficients in the vertical and radial
directions are shown in \Tab{t:nuturb}. In \Fig{fig:Dt_vs_tauf_panel} we plot
the diffusion coefficients together with the $\alpha$-value based on the
Reynolds stress, $\alpha_{\rm t}$, and the $\alpha$-value based on the Maxwell
stress, $\alpha_{\rm t}^{\rm (mag)}$. We include 1-$\sigma$ fluctuation
intervals on all measurements. The radial diffusion is seen to be much stronger
(around 70\%) than the vertical diffusion. This is also to be expected from
\Fig{fig:turbulence_pars}, since the root-mean-square of the vertical velocity
component is smaller than for the horizontal components, so the velocity
fluctuations in the radial direction are stronger than in the vertical
direction.

From the length of the fluctuation bars in
\Fig{fig:Dt_vs_tauf_panel}, it is clear that the
fluctuations in the turbulent diffusion coefficient are very small for the short friction time limit. Combined with
the fact that the quality of the cosine fit is excellent, this means that the
turbulent transport in that case is well-described as diffusion. For intermediate
friction times, with a freely evolving dust velocity, the fluctuation in the
turbulent diffusion coefficient becomes larger, especially in the radial
direction. The average values of the diffusion coefficients nevertheless stay
approximately constant both for short and intermediate friction times. This
gives confidence in that the short friction time approximation is indeed valid
for very small dust grains.

\subsubsection{Diffusion Prandtl Number}

It is of great interest to compare the measured diffusion coefficients with the
turbulent viscosity, since a popular parametrization of turbulent diffusion is
to set the diffusion coefficient equal to the turbulent viscosity coefficient.
It is seen from \Tab{t:nuturb} that the vertical diffusion coefficient is generally around a factor of three to
four times the non-magnetic $\alpha$-value, but the value is comparable to the
magnetic $\alpha$-value. The radial diffusion coefficient is slightly higher
than the total turbulent $\alpha$-value.

We quantify the difference between the measured turbulent diffusion
coefficients and the turbulent viscosity through the diffusion Prandtl number
${\rm Pd}$. This is defined as the ratio between the turbulent viscosity and
the turbulent diffusion coefficient as
\begin{equation}
  {\rm Pd} = \frac{\nu_{\rm t}}{D_{\rm t}} \, .
\end{equation}
For anisotropic turbulence, the Prandtl number depends on the direction.
Unfortunately there is no way to estimate the turbulent viscosity in the
vertical direction, as there is no background shear and thus no flux of
angular momentum vertically, so we shall use the value for the radial turbulent viscosity
even for the vertical Prandtl number.

The measured Prandtl numbers are shown in the last
two columns of \Tab{t:nuturb}. The vertical diffusion Prandtl number is found
to be above unity in the range ${\rm Pd}_z=1.27\ldots1.60$, while the radial
diffusion Prandtl number is below unity in the range ${\rm
Pd}_x=0.79\ldots0.90$ and falling with increasing resolution. This is quite
surprising as Prandtl numbers smaller than one could not be expected from
standard diffusion theory. It is not possible to say whether the vertical
Prandtl number would be similarly low if we had scaled with the proper vertical
turbulent viscosity, because this quantity is, as mentioned above, not known.

\subsubsection{Dependence on Particle Size}

Much analytical work has been
devoted to parametrizing the dependence of the diffusion coefficient on dust
particle radius
\citep{Safronov1969,Voelk+etal1980,Cuzzi+etal1993,Dubrulle+etal1995,SchraeplerHenning2004,Reeks2005}.
According to \cite{SchraeplerHenning2004}, ignoring the effect of the mean
motion of the dust grains, the diffusion coefficient can be written as
\begin{equation}
  D_{\rm t} = \frac{D_0}{1+{\rm St}} \, .
\end{equation}
Here ${\rm St}$ is the Stokes number, and the factor $1/(1+{\rm St})$
determines the variation of diffusion coefficient with particle radius. The
Stokes number is defined as the ratio of the friction time to the turn-over
time $\tau_{\rm c}$ of the largest eddies. Assuming that the rotation speed of
the largest eddies as $v_{\rm e}=\alpha_{\rm t}^q c_{\rm s}$ and choosing $q=0.5$,
one can \citep[following][]{SchraeplerHenning2004} derive the expression
\begin{equation}
  D_{\rm t} = \frac{D_0}{1+4^{-1}\pi\varOmega_0 \tau_{\rm f}} \, .
\end{equation}
Thus for the largest grains considered in this work, with $\varOmega_0 \tau_{\rm
f}=0.02$, the expected change in diffusion coefficient due to particle size is
around $1.5\%$. This is well below the fluctuation intervals in the
measurements. Our results indeed confirm that there is no apparent
size-dependence on the measured diffusion coefficient for our chosen grain
size range. This also confirms the
interpretation that the variation in the observed disc thickness at various
wavelengths is due to differential settling between particles of different
sizes \citep[e.g.][]{DullemondDominik2004} and not due to a variation in the
diffusion coefficient with particle size.

\subsection{Local Dust Density Enhancement}

\begin{figure}
  \includegraphics[width=8.3cm]{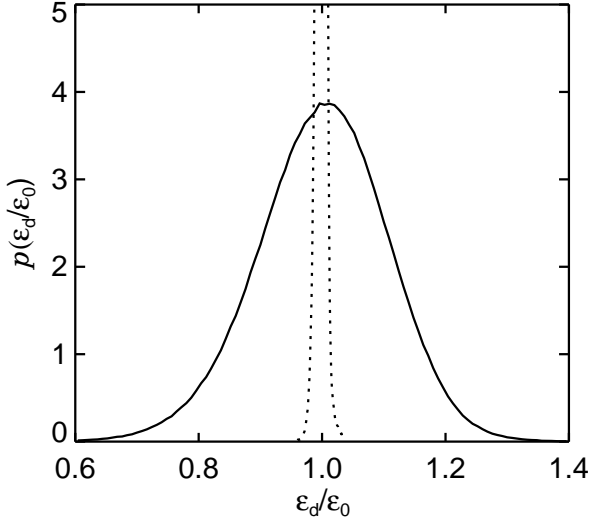}
  \caption{Dust-to-gas ratio probability distribution function for runs with
  $\varOmega_0 \tau_{\rm f}=0.02$ (full line) and $\varOmega_0
  \tau_{\rm f}=2\times10^{-7}$ (dotted line) without gravity on the dust. For
  the intermediate friction run, there is a much higher probability for very low
  or very high dust-to-gas ratios, compared to the short friction time run where
  the dust-to-gas ratio is sharply peaked around $\epsilon_{\rm d}=\epsilon_0$.
  The full probability curve for the short friction time run is shown in
  \Fig{fig:pepsd_vs_epsd_shorttauf}.}
  \label{fig:pepsd_vs_epsd}
\end{figure}

\begin{figure}
  \includegraphics[width=8.3cm]{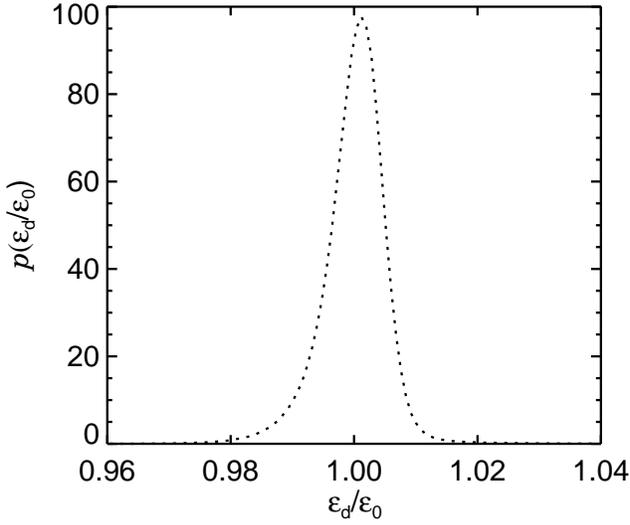}
  \caption{The full dust-to-gas ratio probability distribution function for a
  short friction time run with $\varOmega_0 \tau_{\rm f}=2\times10^{-7}$ and no
  gravity. Because the peak is so sharp compared to the intermediate friction
  time run, only the lower part is shown in \Fig{fig:pepsd_vs_epsd}.}
  \label{fig:pepsd_vs_epsd_shorttauf}
\end{figure}

To explore why the diffusion coefficient fluctuates so much in the intermediate
friction time runs, we plot in \Fig{fig:pepsd_vs_epsd} the dust-to-gas ratio
probability function for an intermediate friction time run with $\varOmega_0
\tau_{\rm f}=0.02$ (full line) and a short friction time run with
$\varOmega_0 \tau_{\rm f}=2\times10^{-7}$ (dotted line) for a resolution of
$64^3$ and no gravity. These two gravity-free runs are named 64c\_ng and
64a\_ng, respectively. The probability of a grid point having a dust-to-gas
ratio between $\epsilon_{\rm d}$ and $\epsilon_{\rm d} + \Delta \epsilon_{\rm
d}$ is
\begin{equation}
  p(\epsilon_{\rm d}) = \frac{\Delta f(\epsilon_{\rm d})}{\Delta
  \epsilon_{\rm d}} \, ,
\end{equation}
where $\Delta f(\epsilon_{\rm d})$ is the fraction of all grid points in the
simulation box having a dust-to-gas ratio between $\epsilon_{\rm d}$ and
$\epsilon_{\rm d} + \Delta \epsilon_{\rm d}$. We average over 10 orbits taken
equidistantly between orbits 10 and 100. According to \Fig{fig:pepsd_vs_epsd}
the probability of finding grid points with very high or very low dust-to-gas
ratios is much higher in the intermediate friction time run than in the short
friction time run. The dust-to-gas ratio in the short friction time run is
extremely peaked around $\epsilon_{\rm d}=\epsilon_0$, thus only the bottom
part of the curve could be shown in \Fig{fig:pepsd_vs_epsd}. The full curve is
shown in \Fig{fig:pepsd_vs_epsd_shorttauf}. 

\begin{figure}
  \includegraphics[width=17.3cm]{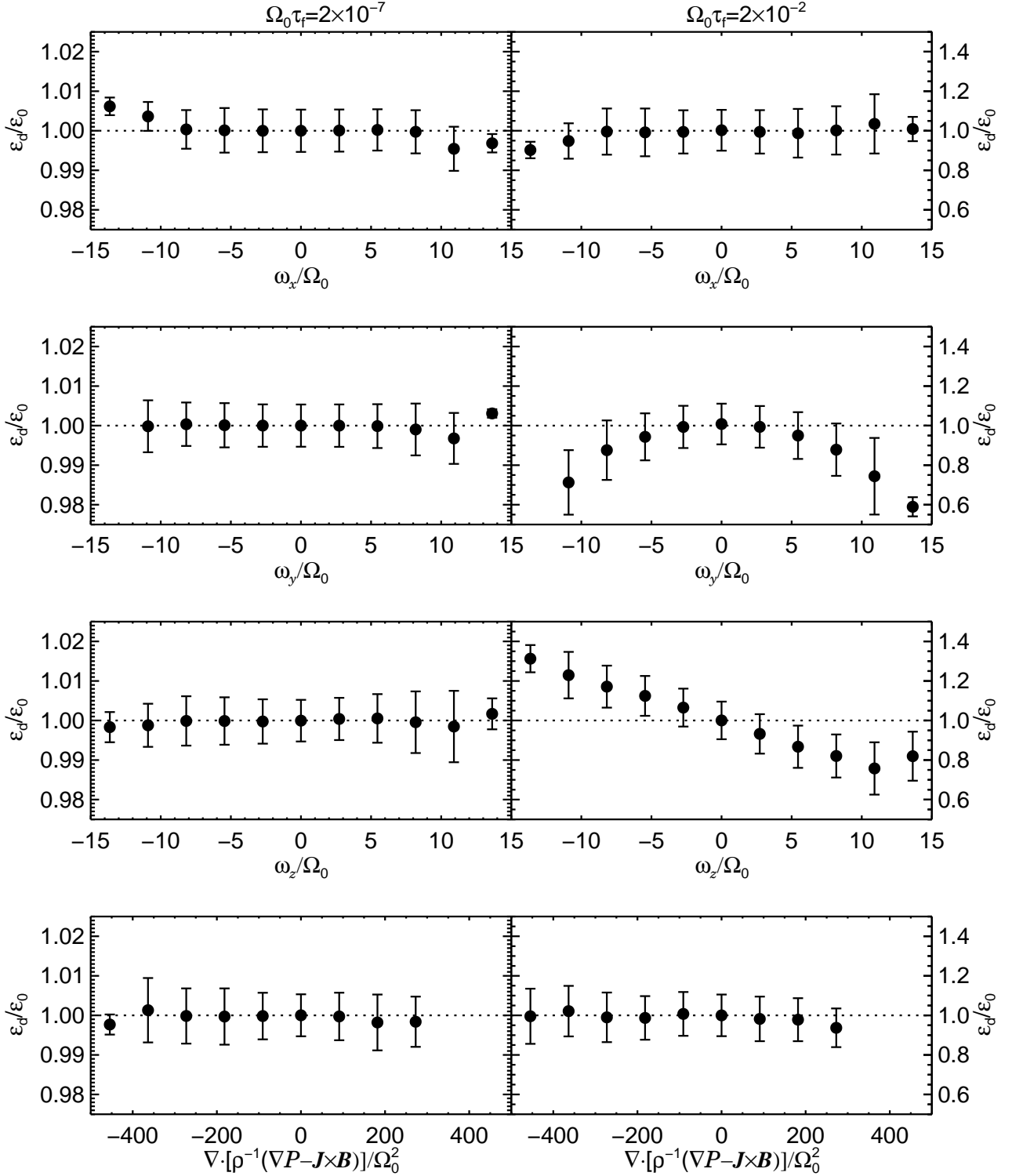}
  \caption{The dust-to-gas ratio $\epsilon_{\rm d}$ in bins of vorticity
  components (first three rows) and divergence of pressure gradient flux (last
  row). The large dot shows the
  average dust-to-gas ratio in the bin, while the bars represent the
  fluctuation interval. The clearest correlation is between vertical vorticity
  and dust-to-gas ratio for the intermediate friction time run. This may be due
  to vortex trapping as explained in the text.} 
  \label{fig:epsd_vs_misc}
\end{figure}

There are two potential sources for the high dust-to-gas ratio contrast that is
seen in the intermediate friction time: trapping of dust grains in turbulent
vortices or trapping in regions of high pressure by pressure gradient trapping,
as mentioned in the introduction. The latter effect can work also in the
short friction time approximation, the first can not.
According to \Eq{eq:w_sfta}, the terminal velocity of small dust grains climbing
up the local pressure gradient is (we ignore gas velocity and set external
gravity to zero)
\begin{equation}\label{eq:w_sfta_nog}
  \vc{w} = \tau_{\rm f}[\rho^{-1}(\nab P - \vc{J}\times\vc{B})] \, .
\end{equation}
The evolution of the dust number density of a fluid element is controlled by
the continuity equation
\begin{equation}\label{eq:dustcont_adv}
  \frac{\De \ln n}{\De t} = -\nab \cdot \vc{w} \, ,
\end{equation}
where $\De/\De t \equiv \dpa/\dpa t + (\vc{w}\cdot\nab)$ is the advective
derivative of the flow. Combining \Eq{eq:dustcont_adv} with \Eq{eq:w_sfta_nog} shows that dust should concentrate in regions
where $\nab \cdot [\rho^{-1}(\nab P - \vc{J}\times\vc{B})] \equiv \nab \cdot
\vc{F} < 0$ and be removed from regions where the divergence is negative. We
examine whether this is the case in the two bottom panels of
\Fig{fig:epsd_vs_misc}. Here the average dust-to-gas ratio (including
1-$\sigma$ fluctuation intervals) is shown for bins in $\nab \cdot \vc{F}$.
The left panel is for the short friction time run 64a\_ng while the right panel
is for the intermediate friction time run 64c\_ng. For the intermediate friction
time run, there is evidently some correlation between a positive divergence and
a low dust-to-gas ratio and vice versa, but the correlation is not very strong.

\begin{figure}
  \includegraphics[width=17.3cm]{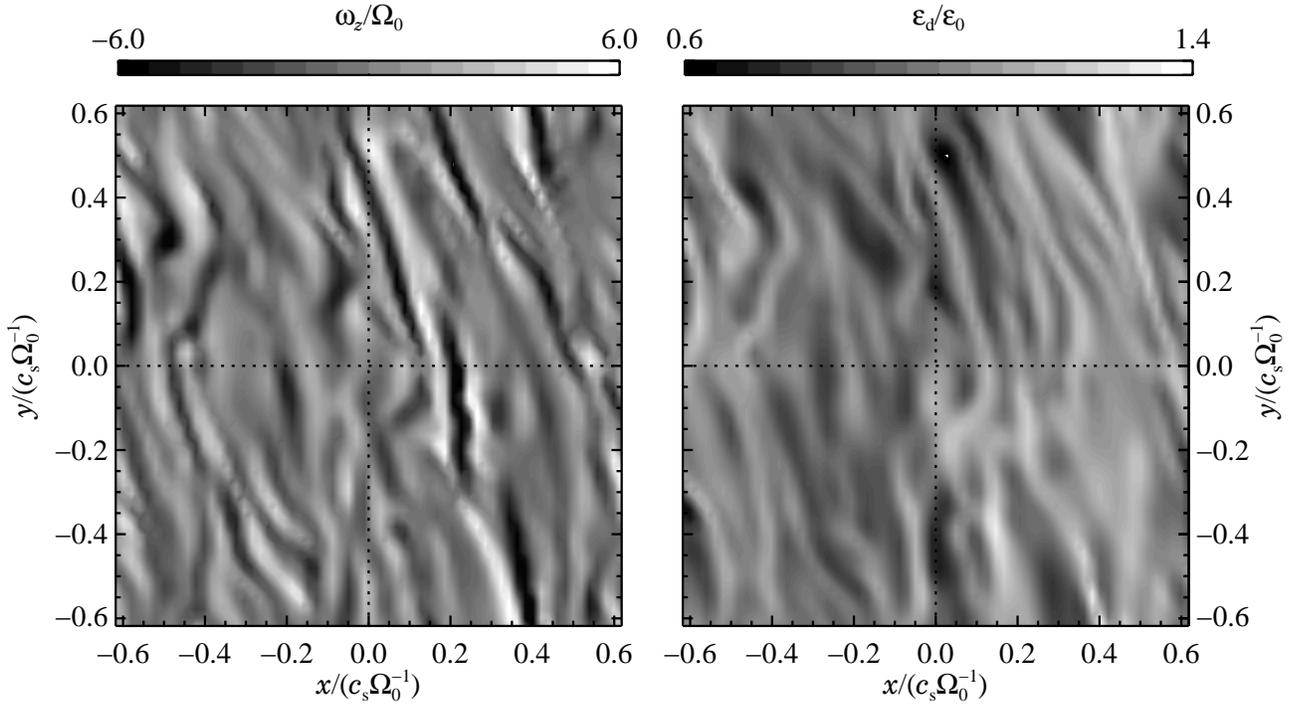}
  \caption{Contour plots of vertical component of vorticity (left panel) and
  dust-to-gas ratio (right panel) in an arbitrary $z$-plane for the
  intermediate friction time run 64c\_ng. There is a tendency for positive
  vorticity (light regions) to correspond to low dust-to-gas ratio (dark
  regions) and vice versa. This indicates that dust grains are being trapped in
  turbulent eddies by the vortex trapping mechanism. The dotted lines are
  reference lines to make comparison between the two plots easier.}
  \label{fig:ooz_epsd}
\end{figure}

Vortex trapping is another potential source of the dust-to-gas ratio contrast
\citep{BargeSommeria1995}. It can be very powerful when $\varOmega_0 \tau_{\rm
f}$ is close to unity. The delayed acceleration of a dust grain entering a
turbulent gas eddy causes the Coriolis force to dominate completely over the
centrifugal force of the eddy. The effect of vortex trapping can be seen from
the vorticity $\vc{\omega}\equiv\nab\times\vc{u}$ of the flow. Cyclonic
vortices (with positive $\omega_z$) have an outwards directed Coriolis force
relative to the center of motion and can expel dust grains. Anticyclonic
vortices (with negative $\omega_z$) have a Coriolis force that points inwards.
Such vortices can trap dust grains. As an illustration of the trapping of dust
grains in turbulent features we show in \Fig{fig:ooz_epsd} contour plots of
$\omega_z$ and $\epsilon_{\rm d}$ in an arbitrarily chosen $x$-$y$-plane. The
vorticity contours show patches of positive and negative vorticity. The
correlation between negative vorticity and high dust-to-gas ratio (and vice
versa) is clearly seen in many places. However, it is not a perfect 1:1 fit, as can also
be expected in a dynamical system that is changing all the time. All
concentrations are only surviving as long as a vortex exists. Turbulent eddies
have a lifetime comparable to the shear time of the system, i.e.\ the orbital
period.

It is easier to see the correlation between vertical vorticity and dust-to-gas
ratio in \Fig{fig:epsd_vs_misc}. Here the three top rows show the correlation
between dust-to-gas ratio and the three directional components of the
vorticity. There is a strong correlation with vertical vorticity component
$\omega_z$ for the intermediate friction time run. This is exactly as expected
in case vortex trapping and expelling is the source of the number density contrast.
A vertical vorticity can however also be caused by a non-rotating flow, e.g.\
if the gas-flow is hyper-Keplerian with a shear velocity that is linear with
the radial coordinate $u_y \propto x$. Such a profile can be caused by a radial
bump in the gas density. Here dust-trapping would be due to pressure gradient
trapping and not due to vortex trapping.

\begin{figure}
  \includegraphics[width=17.3cm]{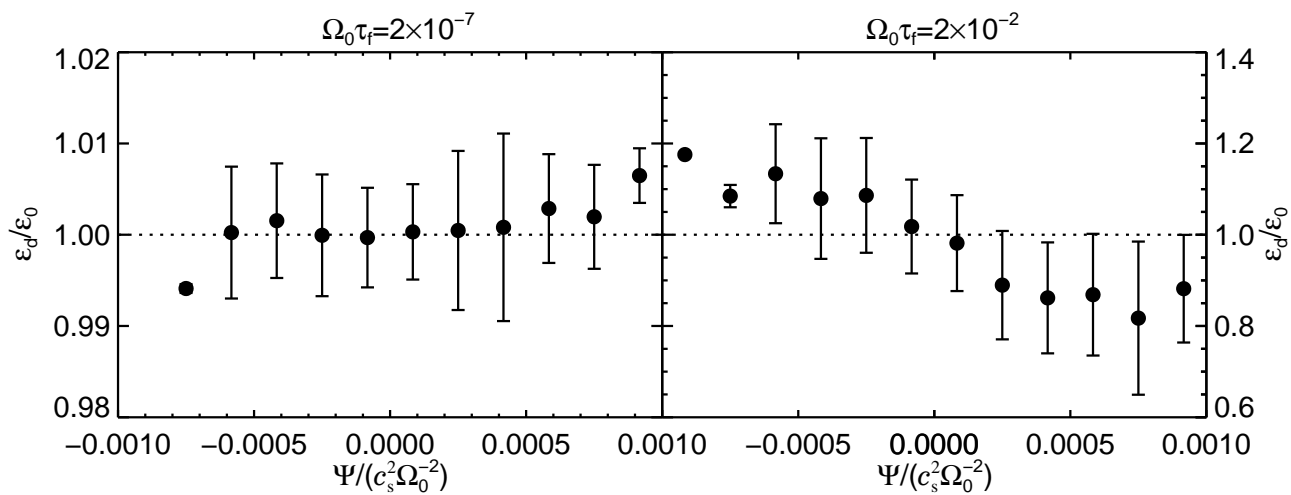}
  \caption{Plot of dust-to-gas ratio in bins of vortex parameter $\varPsi \equiv
  [-(\vc{u}\cdot\nab)\vc{u}] \cdot \vc{f}(\vc{u})$. Anticyclonic vortices have
  a negative value of $\varPsi$, whereas for cyclonic vortices $\varPsi$ is
  positive. For the intermediate friction time run, there is a clear
  anticorrelation between vortex parameter and dust-to-gas ratio. This is an
  indication that dust is being trapped in anticyclonic vortices.}
  \label{fig:epsd_vs_advf}
\end{figure}

A better test of vortex trapping than vertical vorticity can be
devised by taking a closer look at the trapping mechanism \citep[see
e.g.][]{Johansen+etal2004}. An anticyclonic vortex is in equilibrium because
there is a resulting force on the gas particles pointing towards the center of
rotation. This resulting force is a vector sum of the Coriolis force, the
pressure gradient force and the Lorentz force, and it works as a centripetal
force that supplies just the right amount of force necessary to orbit the
center of rotation. In the fluid equations, the resulting centripetal force is
balanced by the additional advection term that keeps the velocity field
unchanged, even though the fluid elements themselves experience an acceleration
towards the center of rotation.
Thus for anticyclonic vortices, the advection vector $-(\vc{u}\cdot\nab)\vc{u}$
points away from the center of rotation, while the Coriolis force
$\vc{f}(\vc{u})$ points towards the center of rotation, which is exactly in the
opposite direction. The occurrence of the Coriolis force pointing in the
opposite direction of the advection vector is a sufficient condition for having
an anticyclonic vortex and thus vortex trapping. For a cyclonic vortex both the Coriolis force and the
advection vector point away from the center of rotation. Defining the vortex
parameter $\varPsi\equiv[-(\vc{u}\cdot\nab)\vc{u}] \cdot \vc{f}(\vc{u})$, we
can now recognize cyclones by a positive value of $\varPsi$ and anticyclones by
a negative value of $\varPsi$. If dust grains are affected by vortex trapping,
then there should be an anticorrelation between $\varPsi$ and the
dust-to-gas ratio at the locations of cyclones and anticyclones. We examine
this in \Fig{fig:epsd_vs_advf}. It is seen that
the anticorrelation between $\varPsi$ and dust-to-gas ratio is 
significant. This allows us to conclude that the large fluctuations in
dust-to-gas ratio for the intermediate friction time runs is caused by
trapping in turbulent eddies.

Curiously there is also a significant correlation between any non-zero toroidal
vorticity component $\omega_y$ and a low dust-to-gas ratio for the intermediate
friction time run in \Fig{fig:epsd_vs_misc}. This may be related to dust grains
being expelled from eddies with a rotation axis parallel to the mid-plane
(in the absence of gravity; when vertical gravity is
included, particles can become suspended in such
eddies, see \citealt{KlahrHenning1997,Pasquero+etal2003}). However, there is no similar correlation with the
radial component of vorticity $\omega_x$, probably because the shear wipes out
any depletions/concentrations on a very short timescale.

A similar search for concentrations of dust grains in MRI turbulence was
performed by \cite{HodgsonBrandenburg1998}. They find that for a frozen gas
velocity field, intermediate friction time dust grains do indeed concentrate in
the turbulent gas structures, but they attribute this effect to dust grains
concentrating where the gas velocity field is converging rather than to vortex
trapping. For an evolving gas velocity field, they find no concentration of
dust. It is not clear why our results differ from these results. However,
\cite{HodgsonBrandenburg1998} focus on concentrations of dust particles in the
vertical plane, while in the current work dust concentrations are most
pronounced in rotating structures in the horizontal plane.

\section{CONCLUSIONS}
\label{ch:conclusions}

The transport properties of dust grains in a turbulent accretion disc is of
interest for many aspects of protoplanetary disc modeling and planet formation
scenarios. In this paper, we have measured the turbulent diffusion coefficient
of dust grains embedded in ideal MHD magnetorotational turbulence directly from
numerical simulations. The choice of magnetorotational turbulence was made
because there is a growing realization that the magnetorotational instability
can work at least in some parts of protoplanetary discs, even where the
ionization fraction may be surprisingly low. It is also routinely produced in
shearing box simulations, so it is a very accessible form of turbulence. Thus,
by the use of MRI, we have a natural source of turbulence, whereas the current
only other alternative for similar studies would be the use of driven
turbulence in a box. 
The use of the ideal MHD
equations can only be justified as a first approach to calculate the turbulent
transport properties of dust grains. Further studies of non-ideal MHD should be
made to clarify the transport properties of grains deeply embedded in the disc
where the ionization fraction is low and where one is confronted with a ``dead
zone'' around the mid-plane of the disc.

As a numerical solver we have used the Pencil Code. This finite difference
code solves the non-conservative form of the dynamical equations. It is
special compared to other codes in that it uses sixth order derivatives in
space. The numerical scheme of the Pencil Code was stabilized using
hyperdiffusivity terms in all the dynamical equations. The effect of
hyperdiffusivity is to affect the large scale motion as little as possible,
while at the same time quenching unstable modes at the smallest scales of the
box. By varying the size of the artificial diffusion coefficient, we have found
the direct influence of artificial diffusion on the measured turbulent
diffusion coefficient to be negligible, most likely due to the fact that mass
diffusion is primarily contributed by the fast and far moving large
scales of the turbulence, and these are as mentioned affected only very little
by hyperdiffusivity. From this perspective, hyperdiffusivity seems to be a tool
that is well suited for measurements of turbulent transport properties.

Since we have only considered dust grains of sizes much less than one meter, 
the dust grains could be treated as a fluid interacting with the gas through
a drag force. For the tiniest dust grains, where the friction time is much shorter
than the computational time-step, we have used an algebraic equation to obtain
the dust velocity at each time-step. This short friction time
approximation incorporates the tendency of dust grains to move up the local
pressure gradient of the gas, an effect which we have referred to as pressure gradient
trapping. It can explain such phenomena as the settling of dust grains towards
the mid-plane of a stratified disc and the radial drift of dust grains in discs
with a radial pressure gradient. In the current work, we have also included the
effects of magnetic pressure and tension in the short friction time
approximation. For intermediate friction time dust grains, where the friction
time is within a few orders of magnitude of the orbital period, we have
integrated the dust equation of motion together with the other dynamical
equations. Here acceleration effects are allowed, in the sense that dust grains
are no longer assumed to instantaneously reach a terminal velocity where the 
drag force is balanced by the other forces affecting the dust.
The fact that the time average of the measured diffusion coefficient was
approximately the same for tiny dust grains, using the short friction time
approximation, and intermediate size dust grains, with a free evolution of the
dust velocity, gives some credit to the validity of the short friction time
approximation.

We have chosen to measure the turbulent diffusion coefficient by forcing the
dust grains to settle towards a mid-plane by an external force field. This
settling was eventually balanced by turbulent diffusion away from the
mid-plane. To deduce the value of the turbulent diffusion coefficient, the
equilibrium dust density could then be compared with an analytical solution for
a parametrized diffusion coefficient.
The method works not only for the vertical direction,
but also for the radial direction, so that we have been able to measure both
the vertical and the radial turbulent diffusion coefficients.

For the short friction time runs, the equilibrium dust number
density was excellently fitted with the expected analytical solution. That means
that the turbulent transport of small dust grains is well-described as
diffusion. For intermediate friction times, the equilibrium dust number density
was much more erratic, especially in the radial direction, and did not always
give a good fit. We also found that the dust-to-gas ratio
probability distribution was much wider than in the short friction time runs. To
explore the reason for the large spread in dust-to-gas ratio, we have examined correlations between different
parameters of the gas and the dust-to-gas ratio. A strong
correlation between vertical vorticity component and dust-to-gas ratio was
found. Based on this and an equally strong correlation between 
the sign of the vortex parameter and the dust-to-gas ratio, we conclude that the
spread in dust-to-gas ratio, and thus the fluctuations in the diffusion coefficient, is due to vortex trapping in turbulent eddies
\citep{BargeSommeria1995}. Some weaker indications that pressure
gradient trapping is taking place were also found, but similar to the results of
\cite{Johansen+etal2004}, the over all dominant trapping mechanism is found to be
vortex trapping. The dust-trapping that is seen in the current work happens for
relatively well-coupled particles with a friction time on the order of a few
percent times the shear time $\varOmega_0^{-1}$. One can speculate that for larger particles,
dust-trapping mechanisms will be so efficient that the diffusion picture of
turbulent transport will no longer be valid, but further investigations into
the transport of larger particles will have to examine this. Such an
investigation would have to incorporate dust grains as particles moving on top
of the gas fluid, since the fluid description of dust grains is no longer valid
when the mean free path becomes larger than the scale height of the disc.

In the vertical direction the turbulent diffusion coefficient was measured to
be smaller than the total turbulent viscosity and have a 
diffusion Prandtl number of approximately ${\rm Pd}_z=1.5$. The diffusion
coefficient is still considerably larger than the non-magnetic turbulent
viscosity alone. The measured radial turbulent diffusion coefficient turned out
to be almost twice as large as the vertical diffusion coefficient. It is
systematically larger than the total turbulent viscosity, i.e.\ the sum of the
non-magnetic and the magnetic turbulent viscosity, with a diffusion Prandtl
number of around ${\rm Pd}_x=0.85$. The value of the radial diffusion Prandtl
number was found to be falling with increasing resolution. Future simulations
should try to find convergence for the diffusion Prandtl numbers, but this is
beyond the scope of the present work.

The anisotropy between the vertical and the radial directions should be taken
into account for studies of planetesimal formation which invoke a gravitational
instability in the dust sublayer. Here the onset of a gravitational instability
in the vertically settled dust layer depends strongly on the effect of vertical
diffusion. The amount of anisotropy can be expected to increase if the effect
of vertical gravity and stratification is included (S. Fromang, personal
communication), since then the buoyancy of the gas would decrease the vertical
velocity fluctuations.

We want to stress that even though we find a radial diffusion Prandtl number of
less than unity, the disc can still be assumed to radially transport dust grains
and angular momentum about equally well. This is important for the modeling of
radial mixing of dust grains and chemical species, a task which is becoming
ever more relevant as more observations of the radial distribution of dust
grains and molecules in protoplanetary discs become available. The result is in agreement
with \cite{Yousef+etal2003}, who find for simulations of forced MHD turbulence a
turbulent magnetic Prandtl number of unity. For dust grains, the equality
between the radial turbulent diffusion coefficient and the turbulent viscosity
is surprising, when considering that most of the angular momentum
is transported by magnetic Maxwell stresses, while the dust grains have no
coupling with magnetic fields at all. Following the argument of
\cite{TennekesLumley1972}, both angular momentum transport by Reynolds stresses
and radial diffusion depend on the radial velocity fluctuations, so one would
expect the non-magnetic $\alpha$-value and the diffusion coefficient to be
similar. The actual cause of the measured mismatch between turbulent diffusion
and non-magnetic turbulent viscosity would seem to need more discussion in the
future.

\acknowledgments

AJ wishes to acknowledge the kind hospitality of Anja Andersen and Axel
Brandenburg at NORDITA in Copenhagen during part of this work. We wish to thank
Cornelis Dullemond for many inspirational discussions on the observational
consequences of dust diffusion. We are also grateful to Coryn Bailer-Jones for
his many suggestions of language improvements to the original manuscript.
Computer simulations were performed at the
Danish Center for Scientific Computing in Odense. Our work is supported in part by the
European Community's Human Potential Programme under contract
HPRN-CT-2002-00308, PLANETS.

\appendix

\section{HYPERDIFFUSIVITY}
\label{ch:hyperdiffusivity}

In this appendix we discuss the use of hyperdiffusivity and present the
hyperversions of viscosity, mass diffusion and resistivity that we are applying
in this work.

Because the Pencil Code is a finite-difference code, artificial diffusivity
terms are needed in all dynamical equations to stabilize the numerical scheme.
For this purpose, we use sixth order hyperdiffusivity terms which affect mainly
high wave numbers, the smallest scales in the simulation, and preserve the
energy at low wave numbers. Hyperviscosity and magnetic hyperdiffusivity have
been used extensively to study the properties of forced magnetohydrodynamical
turbulence \citep[e.g.][and references therein]{BrandenburgSarson2002}. The
prospect is to affect large scales as little as possible by dissipation, thus
widening the inertial range beyond what can be achieved with a regular
viscosity operator.

Possible side effects of using hyperviscosity and magnetic hyperdiffusivity is
to increase the bottleneck effect \citep[a physical effect in turbulence where
energy piles up around the dissipative scale, see e.g.][]{BiskampMueller2000}
and to cause the dynamo-generated magnetic field in helical flows to saturate
at a higher level than what is seen when using a regular viscosity operator
\citep{BrandenburgSarson2002}. Nevertheless, for forced non-magnetic turbulence
\cite{HaugenBrandenburg2004} show that the shape of the inertial range for runs
with hyperviscosity is very similar to the shape for higher resolution runs
without hyperviscosity.

For the current work we define a momentum-conserving hyperviscosity function
$\vc{f}_{\rm \nu}$ as 
\begin{equation}
  \vc{f}_{\rm \nu} (\vc{u},\rho) = 
      (-1)^{m-1} \rho^{-1}\nab\cdot(\nu_m \rho \vc{S}^{(2m-1)}) \, .
\end{equation}
Here $\vc{S}^{(l)}$ is a simplified $l$th order rate-of-strain tensor defined as
\begin{equation}
  S_{ij}^{(l)} = \frac{\dpa^{l} u_i}{\dpa x_j^{l}} \, .
\end{equation}
For computational simplicity we consider the dynamical viscosity $\mu_m \equiv
\nu_m \rho$ to be constant. Then the hyperviscosity function takes the
appearance
\begin{equation}
  \vc{f}_{\rm \nu}(\vc{u},\rho) = (-1)^{m-1} \frac{\mu_m}{\rho} \nabla^{2m} \vc{u} \, ,
  \label{eq:fviscm}
\end{equation}
where $\nabla^{2m} \equiv \nabla_x^{2m} + \nabla_y^{2m} + \nabla_z^{2m}$ is a
high order differential operator that reduces to a Laplacian for $m=1$. For the
purpose of stabilizing the numerical scheme we adopt a sixth order
hyperviscosity by setting $m=3$ in \Eq{eq:fviscm}. The hyperviscosity function
$\vc{f}_\nu $ then appears as
\begin{equation}
  \vc{f}_{\rm \nu}(\vc{u},\rho) = \frac{\mu_3}{\rho} \nabla^6 \vc{u} \, .
  \label{eq:fnu_3rd}
\end{equation}

For the artificial mass diffusion term we define the hyperdiffusion function
$f_{\rm D}$ as
\begin{equation}
  f_{\rm D}(\rho) = (-1)^{m-1} D_m \nabla^{2m} \rho \, ,
\end{equation}
where $D_m$ is a constant diffusion coefficient. Using $f_{\rm D}$ as a
diffusion term in the continuity equation conserves mass density. Again we
adopt a hyperdiffusivity version with $m=3$, leading to the expression
\begin{equation}
  f_{\rm D}(\rho) = D_3 \nabla^6 \rho \, .
  \label{eq:fD_3rd}
\end{equation}

The hyperresistivity function $\vc{f}_\eta$ is defined as
\begin{equation}
  \vc{f}_\eta(\vc{A}) = (-1)^{m-1} \eta_m \nabla^{2 m} \vc{A} \, ,
  \label{eq:fresm}
\end{equation}
where $\eta_m$ is the magnetic diffusivity. As for viscosity and diffusion we
use a hyperresistivity scheme with $m=3$ in \Eq{eq:fresm}. Then the resistivity
function $\vc{f}_\eta$ comes out as
\begin{equation}
  \vc{f}_\eta(\vc{A}) = \eta_3 \nabla^6 \vc{A} \, .
  \label{eq:feta_3rd}
\end{equation}
Using this function as a resistivity term in the induction equation conserves
all components of the magnetic field $\vc{B}$.

\section{RADIAL DIFFUSION EQUILIBRIUM}
\label{ch:radial_diffusion_eq}

In this appendix we derive the equilibrium dust density when the dust is
exposed to a radial gravity. We define a sinusoidal gravity field similar to
what was done in the $z$-direction as
\begin{equation}
  g_x = -g_0 \sin(k_x x) \, ,
\end{equation}
where $k_x = 2\pi/L_x$ is the radial wave number of the field. In the
horizontal plane the Coriolis force connects the radial and toroidal motions,
so that any velocity in one direction results in an acceleration in the other
direction. If not for the damping effect of friction, dust grains starting with
any non-zero velocity would be forced to move in epicyclic motion. Fortunately
the drag force from the gas permits an equilibrium solution to the dust equation of
motion \Eq{eq:dustmomentumeq}. We solve for $\vc{u}=w_z=0$, $w_x = w_x(x)$ and
$w_y = w_y(x)$ and get the two component equations of the equation of motion
\begin{eqnarray}
  0 &=& -w_x \frac{\dpa w_x}{\dpa x} + 2 \varOmega_0 w_y - 
      \frac{1}{\tau_{\rm f}} w_x - g_0 \sin(k_x x) \, , \\
  0 &=& -w_x \frac{\dpa w_y}{\dpa x} - \frac{1}{2} \varOmega_0 w_x
      - \frac{1}{\tau_{\rm f}} w_y \, .
\end{eqnarray}
Again the advection is ignored. This leads to an algebraic linear system of
equations in $w_x$ and $w_y$,
\begin{eqnarray}
  0 &=& 2 \varOmega_0 w_y 
      - \frac{1}{\tau_{\rm f}} w_x - g_0 \sin(k_x x) \, ,\\
  0 &=& -\frac{1}{2} \varOmega_0 w_x 
      - \frac{1}{\tau_{\rm f}} w_y \, ,
\end{eqnarray}
that has the solution
\begin{equation}
  \left\{
    \begin{tabular}{rcl}
      $w_x$ &$=$& $-\frac{\tau_{\rm f} g_0}{1+\varOmega_0^2\tau_{\rm f}^2}
      \sin(k_x x) \approx -\tau_{\rm f} g_0 \sin(k_x x)$ \\
      $w_y$ &$=$& $\frac{\varOmega_0 \tau_{\rm f}^2 g_0}
          {2(1+\varOmega_0^2\tau_{\rm f}^2)} \sin(k_x x) 
          \approx \frac{1}{2} \varOmega_0 \tau_{\rm f}^2 g_0
      \sin(k_x x)$
    \end{tabular}
  \right. \, .
\end{equation}
Here the approximate expressions are valid to first order in $\varOmega_0
\tau_{\rm f}$. The ratio between toroidal and radial velocity is $|w_y/w_x| =
\frac{1}{2} \varOmega_0 \tau_{\rm f}$, so the toroidally imposed velocity
becomes unimportant with sufficiently short friction time. For this form of
velocity field, the equilibrium continuity equation takes the form
\begin{equation}
  0 = -\frac{\dpa}{\dpa x} 
      \left[ w_x(x) n(x,y) - D_x^{\rm (t)} \frac{\dpa n(x,y)}{\dpa x} \right]
      -\frac{\dpa}{\dpa y} 
      \left[ \{w_y(x)+u_y^{(0)}(x)\} n(x,y) - 
      D_y^{\rm (t)} \frac{\dpa n(x,y)}{\dpa y} \right]
      \, .
      \label{eq:cont_eq_xy}
\end{equation}
By considering solutions to \Eq{eq:cont_eq_xy} of the form $n(x,y)=n(x)$, the
$y$-derivative term of \Eq{eq:cont_eq_xy} vanishes entirely. Then the
equilibrium solution to the continuity equation, when assuming that the total
flux of number density radially through the box is zero, is simply
\begin{equation}
  \ln n = \ln n_1 + \frac{\tau_{\rm f} g_0}{k_x D_x^{\rm (t)}} \cos(k_x x) \, ,
  \label{eq:lnn_x_equi}
\end{equation}
formally identical to the vertical case.




\clearpage


\begin{figure}
  \includegraphics[width=8.3cm]{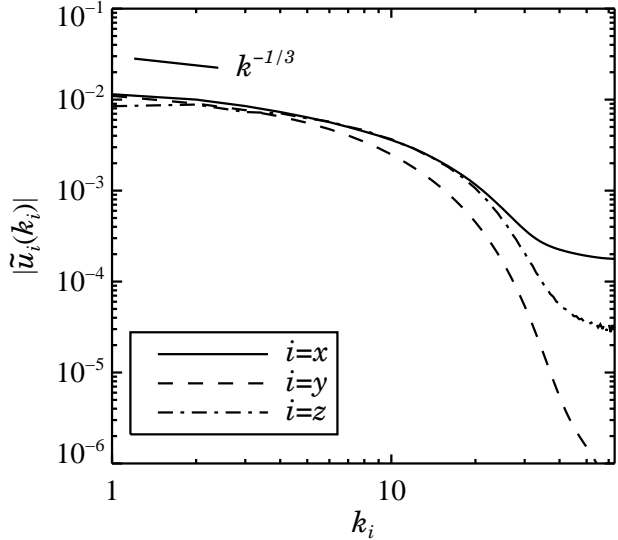}
  \caption{Same as \Fig{fig:power64_uxyz}, but for a $128^3$ resolution run.
  The vertical velocity component still has a flatter slope in the inertial
  range than the two other components, but the surplus power at the very
  smallest scale is greatly diminished compared to the $64^3$ run.}
  \label{fig:power128_uxyz}
\end{figure}






\begin{deluxetable}{cccccccc}

  \tablewidth{0pt}
  \tablecaption{Run parameters}
  \tablehead{\colhead{Run} & \colhead{Res}& \colhead{$g_z^{(0)}$} & 
      \colhead{$g_x^{(0)}$} & \colhead{$\varOmega_0 \tau_{\rm f}$} & 
      \colhead{$a_\bullet/{\rm m}$} & \colhead{SFTA} & 
      \colhead{$\mu_3=\eta_3=D_3$}}

  \startdata

    64a\_z & $64^3$ & $1000$ & $0$ & $2 \times 10^{-7}$ & $2 \times 10^{-7}$ &
          Yes& $1.3 \times 10^{-11}$ \\
    64b\_z & $64^3$ & $10$ & $0$ & $2 \times 10^{-5}$ & $2 \times 10^{-5}$ & 
        Yes& $1.3 \times 10^{-11}$ \\
    64c\_z & $64^3$ & $0.01$ & $0$ & $0.02$ & $0.02$ &
        No& $1.3 \times 10^{-11}$ \\
    128a\_z &$128^3$& $1000 $& $0$ & $2 \times 10^{-7}$ & $2 \times 10^{-7}$ & 
        Yes& $1.3 \times 10^{-12}$ \\
    128c\_z &$128^3$& $0.01$ & $0$ & $0.01$ & $0.01$ &
        No& $1.3 \times 10^{-12}$ \\
    64a\_x & $64^3$ & $0$ & $1000$ & $2 \times 10^{-7}$ & $2 \times 10^{-7}$ &
        Yes& $1.3 \times 10^{-11}$ \\
    64b\_x & $64^3$ & $0$ & $10$ & $2 \times 10^{-5}$ & $2 \times 10^{-5}$ & 
        Yes& $1.3 \times 10^{-11}$ \\
    64c\_x & $64^3$ & $0$ & $0.01$ & $0.02$ & $0.02$ &
        No& $1.3 \times 10^{-11}$ \\
    128a\_x &$128^3$& $0$& $1000$ & $2 \times 10^{-7}$ & $2 \times 10^{-7}$ & 
        Yes& $1.3 \times 10^{-12}$ \\
    128c\_x &$128^3$& $0$ & $0.01$ & $0.01$ & $0.01$ &
        No& $1.3 \times 10^{-12}$ \\
    64a\_ng & $64^3$ & $0$ & $0$ & $2 \times 10^{-7}$ & $2 \times 10^{-7}$ &
        Yes& $1.3 \times 10^{-11}$ \\
    64c\_ng & $64^3$ & $0$ & $0$ & $0.02$ & $0.02$ &
        No& $1.3 \times 10^{-11}$ \\

  \enddata

  \tablecomments{The first column gives the name of the run, the second the
      resolution, the third and fourth the vertical and radial gravity
      strength, the fifth column the friction time, the sixth column the
      corresponding grain radius in a typical solar nebula at
      $r_0=5\,{\rm AU}$, the seventh column whether we used the short friction
      time approximation or not, and the eighth column the value of the
      artificial viscosity $\mu_3$, magnetic diffusivity $\eta_3$ and mass diffusion
      $D_3$.}
  \label{t:rundescrps}
\end{deluxetable}

\begin{deluxetable}{ccccccc}

  \tablewidth{0pt}
  \tablecaption{Turbulent viscosities and turbulent diffusion coefficients}
  \tablehead{\colhead{Run} & \colhead{$\alpha_{\rm t}/10^{-3}$} & 
      \colhead{$\alpha_{\rm t}^{\rm (mag)}/10^{-3}$} &
      \colhead{$D_z^{\rm (t)}/10^{-3}$} &
      \colhead{$D_x^{\rm (t)}/10^{-3}$} &
      \colhead{${\rm Pd}_z$} & \colhead{${\rm Pd}_x$}}

  \startdata
      
    64a\_z  & $0.34 \pm 0.07$ & $1.52 \pm 0.27$ & $1.18 \pm 0.11$ &
        --- & $1.58$ & --- \\
    64b\_z  & $0.34 \pm 0.07$ & $1.52 \pm 0.27$ & $1.18 \pm 0.11$ &
        --- & $1.58$ & --- \\
    64c\_z  & $0.33 \pm 0.08$ & $1.47 \pm 0.29$ & $1.12 \pm 0.14$ &
        --- & $1.60$ & --- \\
    64a\_x  & $0.34 \pm 0.07$ & $1.52 \pm 0.27$ & --- &
        $2.07 \pm 0.28$ & --- & $0.90$ \\
    64b\_x  & $0.34 \pm 0.07$ & $1.52 \pm 0.27$ & --- &
        $2.07 \pm 0.28$ & --- & $0.90$ \\
    64c\_x  & $0.33 \pm 0.08$ & $1.47 \pm 0.29$ & --- &
        $2.12 \pm 0.75$ & --- & $0.85$ \\
    128a\_z & $0.19 \pm 0.03$ & $0.85 \pm 0.12$ & $0.82 \pm 0.10$ &
        --- & $1.27$ & --- \\
    128c\_z & $0.19 \pm 0.04$ & $0.85 \pm 0.19$ & $0.79 \pm 0.13$ &
        --- & $1.31$ & --- \\
    128a\_x & $0.16 \pm 0.02$ & $0.75 \pm 0.10$ & --- &
        $1.15 \pm 0.14$ & --- & $0.79$ \\
    128c\_x & $0.18 \pm 0.03$ & $0.83 \pm 0.12$ & --- &
        $1.27 \pm 0.30$ & --- & $0.79$ \\

  \enddata

  \tablecomments{The first column gives the name of the run. The second and
      third columns show the turbulent $\alpha$-values based on Reynolds and
      Maxwell stresses, respectively. Since there is no back-reaction from the
      dust on the gas, these values are only affected by the dust through the
      dust's
      contribution to the time-step. The next
      two columns show the measured turbulent diffusion coefficients, and in
      the last two columns we write the vertical and radial turbulent diffusion
      Prandtl numbers.}
  \label{t:nuturb}

\end{deluxetable}



\begin{thebibliography}{}

  \bibitem[Arlt \& R\" udiger(2001)]{ArltRudiger2001}
      Arlt, R., \& R\" udiger, G. 2001, \aap, 374, 1035
  \bibitem[Armitage(1998)]{Armitage1998}
      Armitage, P.~J. 1998, \apjl, 501, L189+
  \bibitem[Balbus \& Hawley(1991)]{BalbusHawley1991}
      Balbus, S.~A., \& Hawley, J.~F. 1991, \apj, 376, 214    
  \bibitem[Balbus \& Hawley(1998)]{BalbusHawley1998}
      Balbus, S.~A., \& Hawley, J.~F. 1998, Reviews of Modern Physics, 70, 1
  \bibitem[Barge \& Sommeria(1995)]{BargeSommeria1995}
      Barge, P., \& Sommeria, J. 1995, \aap, 295, L1
  \bibitem[Blum \& Wurm(2000)]{BlumWurm2000}
      Blum, J., \& Wurm, G. 2000, Icarus, 143, 138
  \bibitem[Bockel\' ee-Morvan et~al.(2002)Bockel\' ee-Morvan,
      Gautier, Hersant, Hur\' e, \& Robert]{Bockelee-Morvan+etal2002}
      Bockel\' ee-Morvan, D., Gautier, D., Hersant, F., Hur\' e, J.-M.,
      \& Robert, F. 2002, \aap, 384, 1107
  \bibitem[van Boekel et~al.(2004)van Boekel, Min, Leinert, Waters,
      Richichi, Chesneau, Dominik, Jaffe, Dutrey, Graser, Henning,
      de Jong, K\" ohler, de Koter, Lopez, Malbet, Morel, Paresce,
      Perrin, Preibisch, Przygodda, Sch\" oller, \& Wittkowski]
      {vanBoekel+etal2004}
      van Boekel, R., Min, M., Leinert, C., Waters, L.~B.~F.~M., Richichi,
      A., Chesneau, O., Dominik, C., Jaffe, W., Dutrey, A., Graser, U.,
      Henning, T., de Jong, J., K\" ohler, R., de Koter, A., Lopez, B.,
      Malbet, F., Morel, S., Paresce, F., Perrin, G., Preibisch, T.,
      Przygodda, F., Sch\" oller, M., \& Wittkowski, M.
      2004, \nat, 432, 479
  \bibitem[Boss(2003)]{Boss2003}
      Boss, A.~P. 2003, \apj, 599, 577
  \bibitem[Brandenburg et al.(1995)]{Brandenburg+etal1995} 
      Brandenburg, A., Nordlund, \AA., Stein, R.F., \& Torkelsson, U.
      1995, ApJ, 446, 741
  \bibitem[Brandenburg \& Sarson(2002)]{BrandenburgSarson2002}
      Brandenburg, A., \& Sarson, G.~R.
      2002, Physical Review Letters, 88, 055003
  \bibitem[Brandenburg(2003)]{Brandenburg2003}
      Brandenburg, A.\  2003, in Advances in nonlinear dynamos
      (The Fluid Mechanics of Astrophysics and Geophysics, Vol. 9),
      ed. A. Ferriz-Mas \& M. N\' u\~ nez
      (Taylor \& Francis, London and New York), 269-344
  \bibitem[Chavanis(2000)]{Chavanis2000}
      Chavanis, P.~H. 2000, \aap, 356, 1089
  \bibitem[Cuzzi et~al.(1993)Cuzzi, Dobrovolskis, \& Champney]{Cuzzi+etal1993}
      Cuzzi, J.~N., Dobrovolskis, A.~R., \& Champney, J.~M.
      1993, Icarus, 106, 102
  \bibitem[Cuzzi et~al.(2001)Cuzzi, Hogan, Paque, \& Dobrovolskis]{Cuzzi+etal2001}
      Cuzzi, J.~N., Hogan, R.~C., Paque, J.~M., \& Dobrovolskis, A.~R.
      2001, \apj, 546, 496
  \bibitem[Dubrulle et~al.(1995)Dubrulle, Morfill, \& Sterzik]{Dubrulle+etal1995}
      Dubrulle, B., Morfill, G., \& Sterzik, M. 1995, Icarus, 114, 237
  \bibitem[Dubrulle et~al.(2005)Dubrulle, Mari\' e, Normand,
      Richard, Hersant, \& Zahn]{Dubrulle+etal2005}
      Dubrulle, B., Mari\' e, L., Normand, C., Richard, D., Hersant, F.,
      \& Zahn, J.-P.
      2005, \aap, 429, 1
  \bibitem[Dullemond(2002)]{Dullemond2002}
      Dullemond, C.~P. 2002, \aap, 395, 853
  \bibitem[Dullemond \& Dominik(2004)]{DullemondDominik2004}
      Dullemond, C.~P., \& Dominik, C. 2004, \aap, 421, 1075
  \bibitem[Fleming \& Stone(2003)]{FlemingStone2003}
      Fleming, T., \& Stone, J.~M. 2003, \apj, 585, 908
  \bibitem[Fromang et~al.(2002)Fromang, Terquem, \& Balbus]{Fromang+etal2002}
      Fromang, S., Terquem, C., \& Balbus, S.~A. 2002, \mnras, 329, 18
  \bibitem[Goldreich \& Tremaine(1978)]{GoldreichTremaine1978} 
      Goldreich, P. \& Tremaine, S. 1978, ApJ, 222, 850
  \bibitem[Hanner(1999)]{Hanner1999}
      Hanner, M.~S. 1999, Space Science Reviews, 90, 99
  \bibitem[Haugen et~al.(2004)Haugen, Brandenburg, \& Dobler]{Haugen+etal2004}
      Haugen, N.~E., Brandenburg, A., \& Dobler, W. 2004, \pre, 70, 016308
  \bibitem[Haugen \& Brandenburg(2004)]{HaugenBrandenburg2004}
      Haugen, N.~E., \& Brandenburg, A. 2004, \pre, 70, 026405
  \bibitem[Hawley et~al.(1995)Hawley, Gammie, \& Balbus]{Hawley+etal1995}
      Hawley, J.~F., Gammie, C.~F., \& Balbus, S.~A. 1995, \apj, 440, 742
  \bibitem[Hodgson \& Brandenburg(1998)]{HodgsonBrandenburg1998}
      Hodgson, L.~S., \& Brandenburg, A. 1998, \aap, 330, 1169
  \bibitem[Ilgner et~al.(2004)Ilgner, Henning, Markwick, \& Millar]{Ilgner+etal2004}
      Ilgner, M., Henning, T., Markwick, A.~J., \& Millar, T.~J.
      2004, \aap, 415, 643
  \bibitem[Johansen et~al.(2004)Johansen, Andersen, \& Brandenburg]{Johansen+etal2004}
      Johansen, A., Andersen, A.~C., \& Brandenburg, A. 2004, \aap, 417, 361    
  \bibitem[Klahr \& Henning(1997)]{KlahrHenning1997}
      Klahr, H.~H., \& Henning, T. 1997, Icarus, 128, 213
  \bibitem[Klahr \& Bodenheimer(2003)]{KlahrBodenheimer2003}
      Klahr, H.~H., \& Bodenheimer, P. 2003, \apj, 582, 869    
  \bibitem[Klahr(2004)]{Klahr2004}
      Klahr, H. 2004, \apj, 606, 1070
  \bibitem[Klahr \& Bodenheimer(2005)]{KlahrBodenheimer2005}
      Klahr, H.~H., \& Bodenheimer, P. 2005, submitted to \apj
  \bibitem[Li et~al.(2000)Li, Finn, Lovelace, \& Colgate]{Li+etal2000}
      Li, H., Finn, J.~M., Lovelace, R.~V.~E., \& Colgate, S.~A.
      2000, \apj, 533, 1023   
  \bibitem[Lin \& Papaloizou(1980)]{LinPapaloizou1980}
      Lin, D.~N.~C., \& Papaloizou, J. 1980, \mnras, 191, 37
  \bibitem[M\" uller \& Biskamp(2000)]{BiskampMueller2000}
      M\" uller, W., \& Biskamp, D. 2000, Physical Review Letters, 84, 475
  \bibitem[Norton(2002)]{Norton2002}
      Norton, O.~R. 2002, The Cambridge Encyclopedia of Meteorites
      (Cambridge University Press, 2002)
  \bibitem[Pasquero et~al.(2003)Pasquero, Provenzale, \& Spiegel]{Pasquero+etal2003}
      Pasquero, C., Provenzale, A., \& Spiegel, E.~A. 2003, Physical Review
      Letters, 91, 054502
  \bibitem[Reeks(2005)]{Reeks2005}
      Reeks, M.~W. 2005, Journal of Fluid Mechanics, 522, 263
  \bibitem[Ryu \& Goodman(1992)]{RyuGoodman1992}
      Ryu, D., \& Goodman, J. 1992, \apj, 388, 438
  \bibitem[Safronov(1969)]{Safronov1969}
      Safronov, V.~S. 1969, Evoliutsiia doplanetnogo oblaka.
      (English transl.: Evolution of the Protoplanetary Cloud and Formation of
      Earth and the Planets, NASA Tech. Transl. F-677,
      Jerusalem: Israel Sci. Transl. 1972)    
  \bibitem[Schr\"apler \& Henning(2004)]{SchraeplerHenning2004} 
      Schr\"apler, R. \& Henning, T. 2004, ApJ
  \bibitem[Semenov et~al.(2004)Semenov, Wiebe, \& Henning]{Semenov+etal2004}
      Semenov, D., Wiebe, D., \& Henning, T. 2004, \aap, 417, 93
  \bibitem[Shakura \& Sunyaev(1973)]{ShakuraSunyaev1973}
      Shakura, N.~I., \& Sunyaev, R.~A. 1973, \aap, 24, 337
  \bibitem[Shalybkov \& R\"udiger(2005)]{ShalybkovRuediger2005}
      Shalybkov, D., \& R\"udiger, G. 2005, astro-ph/0501154
  \bibitem[Tennekes \& Lumley(1972)]{TennekesLumley1972}
      Tennekes, H., \& Lumley, J.~L. 1972, First Course in Turbulence
      (Cambridge: MIT Press, 1972)
  \bibitem[V\"olk et~al.(1980)V\"olk, Morfill, Roeser, \& Jones]{Voelk+etal1980}
      V\"olk, H.~J., Morfill, G.~E., Roeser, S., \& Jones, F.~C. 1980, \aap,
      85, 316
  \bibitem[Weidenschilling(1977)]{Weidenschilling1977}
      Weidenschilling, S.~J. 1977, \mnras, 180, 57
  \bibitem[Yousef et~al.(2003)Yousef, Brandenburg, \& R\"udiger]
      {Yousef+etal2003}
      Yousef, T.~A., Brandenburg, A., \& R\" udiger, G. 2003, \aap, 411, 321
      

\end{thebibliography}
\end{document}